\documentclass{ar-1col}

\usepackage[numbers,sort&compress]{natbib}
\usepackage{latexsym}
\usepackage{amsmath,amsfonts,amssymb,amscd,euscript,amstext,amsbsy,,amsthm}

\newcommand{\bea}{\begin{eqnarray}}
\newcommand{\eea}{\end{eqnarray}}
\newcommand{\beq}{\begin{equation}}
\newcommand{\eeq}{\end{equation}}
\newcommand{\nn}{\nonumber}
\usepackage{graphicx}
\usepackage{color}

\usepackage[normalem]{ulem}
\renewcommand\sout{\bgroup \color[rgb]{0.55,0.00,0.99} \ULdepth=-.5ex \ULset}

\begin{document}

\markboth{Barbara Pasquini and Marc Vanderhaeghen}{Dispersion Relations}

\title{Dispersion Theory in Electromagnetic Interactions\footnote{Posted with permission from the {\it Annual Review of Nuclear
and Particle Science}, Volume 68 \copyright~2018 by Annual Reviews, http://www.annualreviews.org.}}

\author{Barbara Pasquini,$^{1,2}$  and \\ Marc Vanderhaeghen$^{3,4}$
\affil{$^1$ Dipartimento di Fisica, Universit\`a degli Studi di  Pavia, Pavia, Italy, 27100; email: barbara.pasquini@unipv.it}
\affil{$^2$ Istituto Nazionale di Fisica Nucleare, Sezione di Pavia, Pavia, Italy, 27100}
\affil{$^3$ Institut f\"ur Kernphysik, Johannes Gutenberg Universit\"at,  Mainz, Germany, D-55099}
\affil{$^4$ PRISMA Cluster of Excellence, Johannes Gutenberg-Universit\"at,  Mainz, Germany, D-55099}}
\begin{abstract}
We review various applications of dispersion relations (DRs) to the electromagnetic structure of hadrons. 
We discuss the way DRs allow one to  extract information on hadron structure constants by connecting information from complementary scattering processes.
We consider the real and virtual Compton scattering processes off the proton, and summarize  recent advances in the  DR analysis of  experimental data to extract the
proton polarizabilities, in comparison with alternative studies based on chiral effective field theories.
We discuss a multipole analysis of real Compton scattering data, along with a DR fit of the energy-dependent dynamical polarizabilities.
Furthermore, we review  new sum rules for the double-virtual Compton scattering
process off the proton, which allow for model independent relations between polarizabilities in real and virtual Compton scattering, and moments of nucleon structure functions. The information on the double-virtual Compton scattering is used to predict and constrain the polarizability corrections to muonic hydrogen spectroscopy.
\end{abstract}

\maketitle

\tableofcontents

\let\clearpage\relax
\allowdisplaybreaks

\section{Introduction}

The starting point of dispersion relations (DRs) dates back to 1926-1927 with the historic papers of Kronig~\cite{Kronig} and Kramers~\cite{Kramers}, discussing the classical dispersion 
of light and the relation between the real and imaginary parts of the index of refraction. They emphasized that a specific relation between the real (dispersive) and imaginary (absorptive) part of the index of refraction was based on the fundamental requirement of causality, in addition to the usual conditions on the scattering matrix, namely, unitarity and Lorentz invariance.
The quantum mechanical formulation of the causality condition was then used in the work  by Gell-Mann, Goldberger and Thirring~\cite{GGT} to derive DRs for  forward Compton 
scattering, using perturbation theory for the electromagnetic interaction. 
Soon after, Goldberger~\cite{Goldberger}  posed the proof on more general grounds, going beyond the limitation of perturbation theory.
These pioneering works laid the foundations for the derivation of a number of sum rules, obtained by combining DRs and low energy theorems~\cite{GG-LET,Low:1954kd,Abarbanel,Klein} for the forward real Compton scattering (RCS) amplitude. 
The best known sum rules are  the Baldin sum rule~\cite{Baldin} for the sum of the dipole polarizabilities 
and the
Gerasimov-Drell-Hearn (GDH) sum rule~\cite{Gerasimov:1965et,Drell:1966jv} for the anomalous magnetic moment.
Further relations can be obtained considering higher-order terms in the low-energy expansion (LEX) of the RCS forward amplitudes.
These sum rules  all relate a measured electromagnetic structure quantity to an
integral over a photo-absorption cross section on the nucleon, and are thus model-independent relations.
The photo-absorption cross sections are by now fairly well known, and have been used in various phenomenological works for the evaluation of the forward RCS sum rules, as reviewed in Section~\ref{FRCS}.

Along with the study of  DRs for forward RCS, in the 1960s there was  considerable work to extend the general formalism of DRs to non-forward RCS, see, for example, Refs.~\cite{Akiba,Lapidus-2,Holliday-1}.
However, DRs for non-forward RCS have become a practicable tool for a detailed investigation of nucleon structure only recently thanks to the advent of high-precision experiments with electromagnetic probes.
Among the most successful applications is the analysis of RCS observables from low energies up to the $\Delta(1232)$-resonance region to extract information on the nucleon static polarizabilities~\cite{Lvov:1996rmi,Babusci:1998ww,Drechsel:1999rf,Drechsel:2002ar,Schumacher:2005an,Pasquini:2007hf}. The static polarizabilities are nucleon structure constants which measure the global strength of the induced current and magnetization densities in the nucleon under the influence of an external quasi-static electromagnetic field. Polarizabilities acquire an energy dependence due to internal
relaxation mechanisms, resonances and particle production thresholds in a physical system. This energy
dependence defines the dynamical polarizabilities~\cite{Babusci:1998ww}, which parametrize the response of the internal degrees of freedom of a composite object to an external, real photon field of arbitrary energy.
Recent advances in the extraction of both the static and dynamical polarizabilities within different variants of DR techniques are summarized in Sections~\ref{RCS-st-pol} and \ref{RCS-dyn-pol}.

When considering the virtual Compton scattering (VCS) process, where   the incident real photon is replaced
by a virtual photon, we can get access to generalized polarizabilities (GPs)~\cite{Guichon:1995pu,Guichon:1998xv}. They depend on the  virtuality of the incident photon
  and allow us to map out the spatial distribution of the polarization densities in a target. The DR formalism for VCS on a proton has been developed more recently~\cite{Pasquini:2000pk,Pasquini:2001yy}, and 
  applied  to a new generation of VCS experiments to extract the scalar GPs of the proton. The state-of-the-art of the dispersion analysis for VCS is presented in Section~\ref{VCS}.
    
The most general case of a double-virtual Compton process, with both initial and final virtual photons, has up to now only been studied in some special limits. The most useful extension is given by the forward double-virtual Compton (VVCS) process, where the initial and final photons have the same non-zero spacelike virtuality. 
In contrast to the processes discussed above, the forward VVCS process is not directly measurable. However, DRs provide a powerful tool to reconstruct the VVCS amplitudes from the empirical information on the electro-absorption cross sections~\cite{Kuhn:2008sy,Chen:2010qc}, encoded in the nucleon structure functions
This is possible provided the integrals converge, otherwise a subtraction is required. One can thus formulate extensions of the Baldin, GDH, and other sum rules, through moments of nucleon structure functions~\cite{Burkhardt:1970ti,Anselmino:1988hn,Ji:1999mr,Drechsel:2002ar}. Such relations can be tested provided one can rely on a theory, such as chiral effective field theory, to calculate the coefficients in the LEXs of the VVCS amplitudes. Recently, several new sum rules have been developed which yield  model-independent relations between polarizabilities in RCS, VCS, and moments of nucleon structure functions. We review the status of the field of VVCS in 
 Section~\ref{VVCS}. 
In Section~\ref{TPE}, we then discuss how the VVCS amplitudes enter to predict and constrain the polarizability corrections to muonic hydrogen spectroscopy. 

We conclude this review in Section~\ref{outlook} and outline some remaining issues for future work.


\section{Real Compton scattering}
\label{RCS}

In this section, we introduce the sum rules for forward RCS along with a review of the most recent  evaluations of the corresponding dispersion integrals.
We then discuss the general formalism  of DRs for non-forward Compton scattering and its application to the extraction of the static and dynamical polarizabilities.

\subsection{Forward dispersion relations} 
\label{FRCS}

Let us consider the kinematics of the RCS reaction, i.e. 
\begin{equation}
\gamma(q) + N(p)\rightarrow \gamma(q')+N(p'),
\end{equation} 
where the variables in brackets denote the four-momenta of the participating particles. 
The initial and final photons are characterized by the polarization four-vectors $\epsilon_\lambda=(0,\boldsymbol{\epsilon}_\lambda)$ and $\epsilon'_{\lambda'}=(0,\boldsymbol{\epsilon}_{\lambda'})$, respectively.
The familiar Mandelstam variables are 
\begin{eqnarray}
s=(p+q)^2,\quad u=(q-p')^2,\quad t=(q-q')^2,
\end{eqnarray}
which are constrained by $s+t+u=2M^2$, with $M$ the nucleon mass.
To describe
Compton scattering, we can choose  the two Lorentz invariant variables $t$ and $\nu$, with the 
the crossing symmetric variable $\nu$ defined by
$
\nu=(s-u)/4M.
$
They are related to the initial ($E_\gamma$) and final ($E'_\gamma$) photon lab energy and to the scattering angle $\theta_{{\rm lab}}$ by
\begin{eqnarray}
\nu=E_\gamma+\frac{t}{4M}=\frac{1}{2}(E_\gamma+E'_\gamma),
\qquad t=-4E_\gamma \, E'_\gamma \, \sin^2 (\theta_{\rm lab}/2)=
-2M(E_\gamma-E'_\gamma).\nn
\end{eqnarray}
In the forward direction,  $\nu=E_\gamma=E'_\gamma$ coincides with the initial and final photon energy and $t=0$.
In this case, the most general form for the  Compton amplitude can be constructed from the independent vectors at our disposal, i.e. $\boldsymbol{\epsilon}$, $\boldsymbol{\epsilon}'$,
$\boldsymbol{q}=\boldsymbol{q}'$, and the proton spin operator $\boldsymbol{\sigma}$, by requiring to be linear in $\boldsymbol{\epsilon}$ and  $\boldsymbol{\epsilon}'$, with the transverse gauge condition $\boldsymbol{\epsilon}\cdot \boldsymbol{q}=\boldsymbol{\epsilon}'\cdot\boldsymbol{q}'=0$, and 
invariant under rotation and parity transformations.
This leads to
\begin{eqnarray}
T(\nu,t=0)={\boldsymbol{\epsilon}}'^{*}_{\lambda'}
\cdot\boldsymbol{\epsilon}_\lambda 
f(\nu)+i\boldsymbol{\sigma}
\cdot(\boldsymbol{\epsilon}'_{\lambda'}{}^*\times
\boldsymbol{\epsilon}_\lambda)g(\nu).
\label{frcs}
\end{eqnarray}
Because of crossing symmetry, the amplitude $T$ also has to be invariant under the transformation $\boldsymbol{\epsilon}\leftrightarrow\boldsymbol{\epsilon}'$
and $\nu\leftrightarrow-\nu$, with the result that $f$ and $g$ are, respectively,  an even and odd function of $\nu$, i.e. $f(- \nu)=f(\nu)$ and $g(- \nu)= - g(\nu)$.
Depending on the relative orientation of the spins, the absorption of the photon leads to hadronic excited states with spin projections $1/2$ and $3/2$.
The optical theorem expresses the unitarity of the scattering matrix by relating the respective cross sections, $\sigma_{1/2}$ and $\sigma_{3/2},$ to the imaginary parts of the forward scattering amplitudes
\bea
{\rm Im}\, f(\nu)&=&\frac{\nu}{8\pi}\left(\sigma_{1/2}(\nu)+\sigma_{3/2}(\nu)\right)\equiv\frac{\nu}{4\pi}\sigma_T(\nu),\nn\\
{\rm Im}\, g(\nu)&=&\frac{\nu}{8\pi}
\left(\sigma_{1/2}(\nu)-\sigma_{3/2}(\nu)\right)
\equiv \frac{\nu}{4\pi}\sigma_{TT}(\nu),
\label{opt-theorem}
\eea
where $\sigma_{T}$ and $\sigma_{TT}$ correspond to the total photo-absorption cross section and to the transverse-transverse interference term, respectively.
Due to the smallness of the fine structure constant, we may neglect all purely electromagnetic processes,
and shall consider only photo-absorption due to the hadronic channels starting at pion production threshold, $\nu_{thr}=m_\pi+m_\pi^2/2M\simeq150$ MeV, where $m_\pi$ is the pion mass.
In order to set up the dispersion integrals, we have to study the behavior of the
absorption cross sections for large energies. The total cross section $\sigma_T$ is essentially constant above the resonance
region, with a slow logarithmic increase at the highest energies, and therefore we must subtract the DR
for $f$. If we subtract at $\nu=0$, we also remove the nucleon-pole terms at this point. 
Using causality, crossing symmetry,  and the optical theorem~\ref{opt-theorem}, the subtracted DR reads:
\bea 
\mathrm{Re}\, f(\nu)&=&f(0)+\frac{\nu^2}{2\pi^2}{\cal P}\int_{\nu_{thr}}^{+\infty} {\rm d}\nu'
\frac{\sigma_T(\nu')}{\nu'^2-\nu^2}.\label{int1}
\eea
For the odd function, we can instead assume the following unsubtracted DR:
\bea 
\mathrm{Re} \, g(\nu)&=&\frac{\nu}{2\pi^2}{\cal P}\int_{\nu_{thr}}^{+\infty} {\rm d}\nu'
\frac{\nu' \sigma_{TT}(\nu')}{\nu'^2-\nu^2}.
\label{int2}
\eea
The behaviour of the scattering amplitudes at low energies is predicted by low-energy theorems (LETs)~\cite{GG-LET,Low:1954kd,Abarbanel,Klein} in the following form:
 \bea
f(\nu)&=&-\frac{e^2e^{2}_{N}}{4\pi M}+(\alpha_{E1}+\beta_{M1})\nu^2+[\alpha_{E1\nu}+\beta_{M1\nu}+\frac{1}{12}(\alpha_{E2}+\beta_{M2})]\nu^4+{\cal O}(\nu^6),\label{taylor-f}\\
g(\nu)&=&-\frac{e^2\kappa_N^2}{8\pi M^2}\nu+\gamma_0\nu^3+\bar{\gamma}_0\nu^5+{\cal O}(\nu^7).\label{taylor-g}
\eea
The leading terms in the expansion of Eqs.~\ref{taylor-f} and \ref{taylor-g} are due to the intermediate nucleon states (Born terms), 
and depend solely on the static properties of the nucleon, i.e. the charge $ee_N$, with $e_p=1$ and $e_n=0$, the mass $M$,
and the anomalous magnetic moment $(e/2M)\kappa_N$, with $\kappa_p=1.79$ and $\kappa_n=-1.91$. 
Only the higher-order terms contain information on the internal structure (spectrum and excitation strengths) of the complex
system. In the case of  the spin-independent amplitude $f(\nu)$, the term ${\cal O}(\nu^2)$ describes Rayleigh scattering and yields information
on the internal nucleon structure through the electric ($\alpha_{E1}$) and magnetic ($\beta_{M1}$) dipole
polarizabilities, while the higher-order terms at ${\cal O}(\nu^4)$  contain contributions of dipole
retardation ($\alpha_{E1\nu}$ and $\beta_{M1\nu}$) and higher multipoles ($\alpha_{E2}$ and $\beta_{M2}$).  
 In the case of the spin-flip amplitude $g(\nu)$, the leading
term is determined by the anomalous magnetic moment, and the higher-order terms  contain information
on the spin structure through the  forward spin polarizability (FSP) $\gamma_0$ and  higher-order FSP
$\bar{\gamma}_0$.
These results can be compared with the coefficients of the Taylor series expansion around $\nu = 0$ of the integrals
 in Eqs.~\ref{int1} and \ref{int2}. In the spin-independent sector, this yields the Baldin sum rule~\cite{Baldin}:
 \bea
 \alpha_{E1}+\beta_{M1}=\frac{1}{2\pi^2}\int_{\nu_{thr}}^{+\infty} {\rm d}\nu' \frac{\sigma_T(\nu')}{\nu'^2}, \label{baldin}
 \eea
and a fourth-order Baldin sum rule for the higher-order static polarizabilities~\cite{Gryniuk:2015eza}:
 \bea
\alpha_{E1\nu}+\beta_{M1\nu}+\frac{1}{12}(\alpha_{E2}+\beta_{M2})=\frac{1}{2\pi^2}\int_{\nu_{thr}}^{+\infty} {\rm d}\nu' \frac{\sigma_T(\nu')}{\nu'^4}.\label{ho-baldin}
 \eea
In the spin dependent sector, the leading term yields the Gerasimov-Drell-Hearn (GDH) sum rule~\cite{Gerasimov:1965et,Drell:1966jv}:
\bea
\frac{\pi e^2\kappa^2_N}{4M^2}=-\int_{\nu_{thr}}^{+\infty} {\rm d}\nu' \frac{\sigma_{TT}(\nu')}{\nu'},\label{GDH}
\eea
while  the higher-order coefficients  gives the Gell-Mann, Goldberger, and Thirring (GGT) sum rule~\cite{GG-LET,GGT} for the FSP $\gamma_0$
and a sum rule for the higher-order FSP $\bar{\gamma}_0$~\cite{Pasquini:2010zr}:
\bea
\gamma_0=\frac{1}{2\pi^2}\int_{\nu_{thr}}^{+\infty} {\rm d}\nu' \frac{\sigma_{TT}(\nu')}{\nu'^3},\label{FSP}\qquad
\bar\gamma_0=\frac{1}{2\pi^2}\int_{\nu_{thr}}^{+\infty} {\rm d}\nu' \frac{\sigma_{TT}(\nu')}{\nu'^5}.
\eea
These dispersive integrals have been evaluated from the available experimental data on the total photo-absorption cross section and helicity-difference photo-absorption cross section, using different 
prescriptions for the extrapolation in the kinematical regions not covered by the data.
The results from the most recent evaluations are collected in  \textbf {Table \ref{tab1}}.
\begin{table}[h]
\tabcolsep7.5pt
\caption{Empirical evaluations of the Baldin sum rule (Eq.~\ref{baldin}), the fourth-order relation in the spin-independent sector (Eq.~\ref{ho-baldin}), the GDH sum rule (Eq.~\ref{GDH}), and the leading- and higher-order FSPs (Eq.~\ref{FSP}).}
\label{tab1}
\begin{center}
\begin{tabular}{@{}l|c|c|c|c|c@{}}
\hline
 &Baldin&IV order&GDH &$\gamma_0$ &$\bar{\gamma}_0$ \\
 &($10^{-4}$ fm$^3$)& ($10^{-4}$ fm$^5$) & $(\mu$b)& ($10^{-6}$ fm$^4$) &($10^{-6}$ fm$^6$)\\
\hline
Babusci et al.~\cite{Babusci:1997ij} & $13.69\pm 0.14$  &  & &&\\
A2~\cite{OlmosdeLeon:2001zn} & $13.8\pm 0.4$  & & &&\\
Gryniuk et al.~\cite{Gryniuk:2015eza} & $14.0\pm 0.2$ &$6.04\pm0.4$ & & &\\
GDH  \& A2~\cite{Ahrens:2001qt,Dutz:2003mm,Helbing:2006zp} & & &$212\pm 17.1$
&$-101\pm 13$\\
Pasquini et al.~\cite{Pasquini:2010zr}& & &$210\pm 15.2$
&$-90\pm 14$
&$ 60\pm 10$\\
Gryniuk et al.~\cite{Gryniuk:2016gnm} & & &$204.5\pm 21.4$ &$-92.9\pm 10.5$&$48.4\pm8.2$\\
sum rule &&&204.78&\\
\hline
\end{tabular}
\end{center}
\end{table}

The database for  the total photo-absorption cross section covers  the energy intervals [0.2, 4.2] GeV~\cite{Armstrong:1971ns,Bartalini:2008zza,Sandorfi}  and [18, 185] GeV~\cite{Caldwell:1978yb}, with additional two measurements at 200 GeV~\cite{Aid:1995bz} and 209 GeV~\cite{Chekanov:2001gw}. The contribution to the sum rules from $\nu_{thr}$ to 0.2 GeV  was determined on the basis of various multipole analyses for pion photo-production, i.e. MAID~\cite{Drechsel:1998hk,Drechsel:2007if}, SAID~\cite{Workman:2011vb} and HDT~\cite{Hanstein:1997tp}, while the contribution from the region above 2 GeV, extrapolated to $+\infty$, was obtained from different fits, mainly based on Regge theory.

The evaluations of the sum rules in the spin-dependent sector are mainly based on the recent  GDH-Collaboration data for the helicity-difference photo-absorption cross section, covering the region from 0.2 to 2.9 GeV~\cite{Ahrens:2000bc,Ahrens:2001qt,Dutz:2003mm}.
In Ref.~\cite{Pasquini:2010zr}, the data set has been supplemented with measurements of the polarized differential cross sections 
for the $n\pi^+$ channel up to energy equal to $0.175$ GeV. These data points have been extrapolated  into the unmeasured angular range with the HDT analysis, in order  to reconstruct the helicity-difference total cross section.
Considering the  energy-weighting factors in the dispersion integrals of Eqs.~\ref{GDH} and \ref{FSP}, one finds that the most crucial contribution for the FSP and higher-order FSP is from the threshold  region up to the first resonance region. 
In particular,  the contribution from the charged-pion channel is  characterized by a strong competition between the $E_{0+}$ multipole above threshold and the $M_{1+}$ near the $\Delta(1232)$ resonance, whereas the neutral-pion channel is almost completely described by the $\Delta(1232)$ resonance effects.
As a result, the evaluation of the sum rules for the FSPs can be particularly sensitive to the multipole analysis used for the extrapolation of the integrand into the unmeasured region near threshold.
The final results, as summarized in \textbf {Table \ref{tab1}}, are all consistent, within the error bars, and reproduce the GDH sum rule value ({\it lhs} of Eq.~\ref{GDH}).

\subsection{Static polarizabilities}
\label{RCS-st-pol}
The physical content of the static polarizabilities can be best illustrated using  effective multipole interactions for the coupling of the electric ($\boldsymbol{E}$) and magnetic ($\boldsymbol{H}$) fields of a source with the internal structure of the nucleon.
When expanding the
Compton scattering amplitude in the photon energy, 
 the second- and fourth-order contributions read~\cite{Babusci:1998ww,Holstein:1999uu}:
\begin{eqnarray}
H_{\rm{eff}}^{(2)} &=& -4\pi\left[
{\textstyle\frac{1}{2}}\,\alpha_{E1}\,\boldsymbol{E}\,^2 +
{\textstyle\frac{1}{2}}\,\beta_{M1}\,\boldsymbol{H}\,^2\right],
\label{h2order}
\\
H_{\rm{eff}}^{(4)} & = &-4\pi\left[
{\textstyle\frac{1}{2}}\,\alpha_{E1\nu}\,\dot{\boldsymbol{E}}\,^2 +
{\textstyle\frac{1}{2}}\,\beta_{M1\nu}\,\dot{\boldsymbol{H}}\,^2
+ {\textstyle\frac{1}{12}}\,\alpha_{E2}\,E_{ij}^2+
{\textstyle\frac{1}{12}}\,\beta_{M2}\,H_{ij}^2 \right],
\label{h4order}
\end{eqnarray}
where the dots denote a time derivative, and  the quadrupole field tensors are denoted by:
\bea
E_{ij}=\frac{1}{2}\left(\nabla_i E_j+\nabla_i E_i\right),\qquad H_{ij}=\frac{1}{2}\left(\nabla_i H_j+\nabla_i H_i\right).
\eea
In Eq.~\ref{h2order} we recognize the static electric ($\alpha_{E1}$) and magnetic ($\beta_{M1}$) polarizabilities, describing the dipole deformations of the electric and magnetic densities inside the nucleon induced by external static electromagnetic fields.
At higher order, the terms in $\alpha_{E1\nu}$ and $\beta_{M1\nu}$ are retardation or dispersive corrections to the lowest-order static polarizabilities and describe the response of the system to time-dependent fields. The parameters $\alpha_{E2}$ and $\beta_{M2}$ represent quadrupole polarizabilities
and measure the electric and magnetic quadrupole moments induced in
a system in the presence of an applied field gradient.
The dependence on the spin enters at third-order via the following effective Hamiltonian
\begin{eqnarray}
\label{h3order}
{H}_{\rm{eff}}^{(3)} &=&  - 4 \pi \left[
{\textstyle\frac{1}{2}}\gamma_{E1E1}
\,\boldsymbol{\sigma}\cdot(\boldsymbol{E}\times\dot{\boldsymbol{E}})
+ {\textstyle\frac{1}{2}}\gamma_{M1M1}\,
\boldsymbol{\sigma}\cdot(\boldsymbol{H}\times\dot{\boldsymbol{H}}) 
\right. \nonumber \\
&&\left. 
- \gamma_{M1E2}\, E_{ij}\,\sigma_iH_j
+ \gamma_{E1M2}\, H_{ij}\,\sigma_iE_j  \right],
\end{eqnarray}
where the
four spin  polarizabilities $\gamma_{E1E1}$, $\gamma_{M1M1}$,
$\gamma_{M1E2}$, and $\gamma_{E1M2}$  are related to a multipole
expansion~\cite{Babusci:1998ww}, as reflected in the subscript notation. 
The FSP $\gamma_0$, and the so-called backward spin polarizability $\gamma_\pi$, entering the Compton scattering amplitudes at backward angles, are then obtained through the following combinations:
\bea
\gamma_0=-\gamma_{E1E1}-\gamma_{M1M1}-\gamma_{E1M2}-\gamma_{M1E2},\quad\gamma_\pi=-\gamma_{E1E1}+\gamma_{M1M1}-\gamma_{E1M2}+\gamma_{M1E2}.\quad
\label{gamma0-gammapi}
\eea
The extraction of the polarizabilities from RCS data has become a mature field in recent years.
It is performed mainly by three techniques. The first one
is a low-energy expansion (LEX) of the RCS cross sections.
Unfortunately this procedure is only applicable at photon energies
well below 100 MeV, which makes a precise extraction
a rather challenging task because of the very low  sensitivity to the polarizabilities at these energies.
This sensitivity is increased by measuring RCS observables
around pion threshold and into the $\Delta(1232)$ region.
A second formalism which has been successfully applied to
RCS data  up to these energies makes use of DRs. It has been worked out for both
unsubtracted~\cite{Lvov:1996rmi} and subtracted~\cite{Drechsel:1999rf,Pasquini:2007hf} DRs. Recently, a third approach has been developed within
the framework of a chiral effective field theory~\cite{Pascalutsa:2002pi,Hildebrandt:2003md,Lensky:2009uv,Lensky:2012ag,McGovern:2012ew,Griesshammer:2012we,Lensky:2015awa},
for energies up to the $\Delta$-resonance region.

In order to set up the DR framework, the first step is to construct
 a complete set of amplitudes  in accordance
with relativity and free of kinematical singularities. According to L'vov et al.~\cite{Lvov:1996rmi}, they can be identified as six Lorentz invariant amplitudes  $A_i(\nu,t)$, $i=1,\dots, 6$, which depend  on the invariants $\nu$ and $t$, and obey the crossing symmetry relation $A_i(\nu,t)=A_i(-\nu,t)$.
Next, causality requires certain analytic
properties of the amplitudes, which allow for a continuation of the scattering amplitudes into the
complex plane and lead to DRs connecting the real and imaginary parts of these
amplitudes. The imaginary parts can be replaced by photo-production amplitudes using
unitarity, and as a result we can complete the Compton amplitudes from experimental
information on photo-absorption  reactions.
In particular, the unsubtracted DRs at fixed $t$ reads:
\begin{equation}
{\rm Re} A_i(\nu, t) = A_i^B(\nu, t) +
{2 \over \pi}  {\mathcal P} \int_{\nu_{thr}}^{+ \infty} d\nu' 
{{\nu'  \mathrm{Im}_s A_i(\nu',t)} \over {\nu'^2 - \nu^2}},
\label{eq:unsub}
\end{equation}
where $A_i^B$ are the nucleon pole contributions  and ${\mathcal P}$ denotes the principal value integral, which runs 
from the pion production threshold upwards.
Taking into account the energy weighting,
 the threshold pion production and the decay
of low-lying resonances yield the largest contributions to
the integral. With existing information on these processes
and reasonable assumptions on the lesser known
higher part of the spectrum, the integrand can be constructed
up to centre-of-mass (cm) energies $W \simeq 2$ GeV.
However, a Regge analysis for the asymptotic behavior does not guarantee
the convergence of the integrals for the amplitudes $A_1$ and $A_2$.
 This behavior is mainly due to fixed poles in the
$t$ channel, notably  the
exchange of a neutral pion for $A_2$ and of a $\sigma$ meson for $A_1$.
  To circumvent this problem, L'vov et al.~\cite{Lvov:1996rmi} proposed to use finite-energy sum rules for these two amplitudes,
  i.e. to close the contour
 integral in the complex plane by a semicircle of finite radius
$\nu_{max}$ and to identify  the contribution from the semicircle  with the asymptotic contribution described by $t$-channel poles.
 This procedure is relatively safe for $A_2$
because the $\pi^0$ pole  is well established
by both experiment and theory. However, it
introduces a considerable model dependence for $A_1$, where the $\sigma$ meson has to be considered as a phenomenological parametrization to model
correlations
in the two-pion scalar-isoscalar channel.

Alternatively,  one can introduce subtracted DRs  to avoid the convergence problem.
A convenient framework has been worked out in Refs.~\cite{Drechsel:1999rf,Pasquini:2007hf}, by suggesting to subtract the fixed-$t$ DRs of Eq.~\ref{eq:unsub} at $\nu=0$, with the result:
\begin{equation}
\mathrm{Re} A_i(\nu, t) = A_i^B(\nu, t) +
\left[ A_i(0, t) - A_i^B(0, t) \right]
+{2 \over \pi} \nu^2 {\mathcal P} \int_{\nu_{thr}}^{+ \infty} d\nu' 
{{ \mathrm{Im}_s A_i(\nu',t)} \over {\nu'  (\nu'^2 - \nu^2)}}.
\label{eq:sub}
\end{equation}
The two extra powers of $\nu'$ in the denominator of the integrand ensure now the convergence of the dispersion integrals for all the amplitudes.
The subtraction functions $A_i(0,t)-A_i^B(0,t)$ in Eq.~\ref{eq:sub} can be determined by once-subtracted DRs in the $t$ channel:
\begin{eqnarray}
A_i(0, t) - A_i^B(0, t) &=&
\left[ A_i(0, 0) - A_i^B(0, 0) \right]
+
\left[ A_i^{t-pole}(0, t) -A_i^{t-pole}(0, 0) \right] \nonumber\\
&+&{t \over \pi}  \int_{4 \, m_\pi^2}^{+ \infty} dt'  {{\mathrm{Im}_t
A_i(0,t')} \over {t' (t' - t)}} +{t \over \pi}  \int_{-
\infty}^{-2m^2_\pi-4Mm_\pi} dt'  {{\mathrm{Im}_t A_i(0,t')} \over {t'  (t'
- t)}} , \label{eq:subt}
\end{eqnarray}
where $A_i^{t-pole}(0, t)$ represents the contribution of the poles in the $t$ channel, in particular of the $\pi^0$ pole in the case of $A_2$ as evaluated in~\cite{Drechsel:1999rf}.
The actual calculation of the dispersion integrals is performed by using the unitarity relation to evaluate the imaginary parts in Eqs.~\ref{eq:sub} and \ref{eq:subt}.
 In the $s$ channel, the unitarity relation is saturated with the $\pi N$ intermediate
states and the resonant contributions of inelastic channels
involving multiple pions. In particular, for the $\gamma N\rightarrow \pi N \rightarrow \gamma N$ channel different analysis of pion-photoproduction, such as MAID~\cite{Drechsel:1998hk,Drechsel:2007if}, SAID~\cite{Workman:2011vb} and the HDT~\cite{Hanstein:1997tp} dispersive analysis, have been employed and compared to control  the uncertainties from this channel to the RCS observables.
The multi-pion intermediate states are approximated by the inelastic decay
channels of the $\pi N$  resonances as detailed in~\cite{Drechsel:1999rf}. 
This simple approximation of the higher inelastic channels is quite
sufficient, because these channels are largely suppressed by
the energy denominator of the subtracted DRs of
Eq.~\ref{eq:sub}.
The imaginary parts in the $t$ channel from $4m_\pi^2\rightarrow +\infty$ are calculated 
using the $\gamma \gamma \rightarrow \pi \pi \rightarrow N \bar N$ channel as input. In a first step, a unitarized
amplitude for the $\gamma \gamma \rightarrow \pi \pi$ subprocess is constructed
from available experimental data. This
information is then combined with the $\gamma\gamma \rightarrow N \bar N$ amplitudes
determined by analytical continuation of $\pi N$ scattering
amplitudes~\cite{Hoehler83}. In
practice, the upper limit of integration along the positive-$t$ cut is  taken equal to $t=0.78$ GeV$^2$,  which is the
highest $t$ value at which the $\pi \pi \rightarrow N\bar N$  amplitudes are tabulated in Ref.~\cite{Hoehler83}.
This serves  well for the present purpose, since the subtracted $t$-channel dispersion integrals converge much below this value.
The second integral in Eq.~\ref{eq:subt} runs along the negative-$t$ cut, from $-\infty$ to $ -2(m^2_\pi+2Mm_\pi)\approx -0.56$ GeV$^2$, and lies in the kinematical unphysical region.
As long as we stay at small (negative) values
of $t$, this integral is strongly suppressed by the denominator
$t'(t'-t)$ in Eq.~\ref{eq:subt}, and can be approximated by taking the analytical continuations at $\nu=0$ and negative $t$ of the most important contributions  from the $\Delta-$resonance and non-resonant $\pi N$ in the physical $s$-channel region.
Having defined the calculation of the $s$- and $t$-channel integrals, we are left with the subtraction constants $a_i= A_i(0, 0)
- A_i^B(0, 0)$
 in Eq.~\ref{eq:subt}, which are directly related to the polarizabilities, as detailed in Ref.~\cite{Drechsel:1999rf}.
\begin{figure}
\vspace{-0.5cm}
\centering
\includegraphics[width=0.49\textwidth]{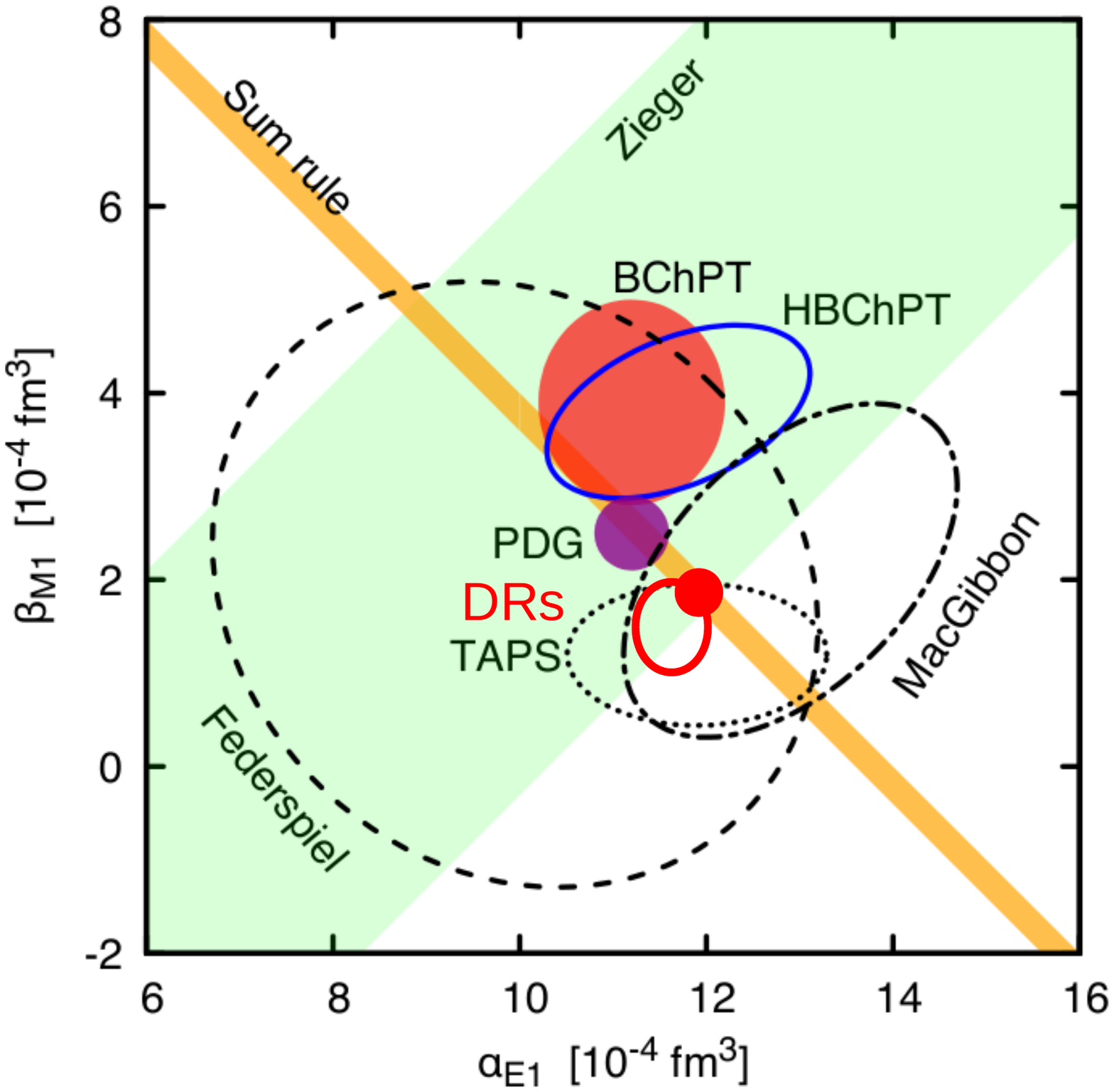}
\vspace{-1.5 truecm}
\caption{The scalar polarizabilities of the proton. The orange band is the average over the Baldin sum rule evaluations listed in  \textbf {Table \ref{tab1}}. The green band shows the experimental constraint on the difference $\alpha_{E1}-\beta_{M1}$ of the dipole polarizabilities from Zieger et al.~\cite{Zieger:1992jq}.  The black curves give the experimental results from Federspiel et al.~\cite{Federspiel:1991yd}, MacGibbon et al.~\cite{MacGibbon:1995in}, and TAPS~\cite{OlmosdeLeon:2001zn}. The  BChPT result is from Ref.~\cite{Lensky:2015awa} and the HBChPT constraint is from Ref.~\cite{McGovern:2012ew}. The red closed and open circles show the results from the fit using subtracted DRs, with and without the constraint of the Baldin sum rule, respectively~\cite{Pasquini:2017ehj}. The plot is adapted from Ref.~\cite{Hagelstein:2015egb}.}
\label{fig1}
\end{figure}
The  optimal strategy would be  to use all these constants as fit parameters to the Compton observables. In practice, 
a simultaneous fit of all the six leading static polarizabilities has not been feasible so-far due to the limited statistics of the available RCS data set.

As a matter of fact, analyses based on subtracted DRs have started to be available only recently, while traditionally the most used analysis tool has been  unsubtracted DRs.
\textbf{Figure \ref{fig1}} shows a summary of the extraction  of the scalar polarizabilities obtained in various frameworks, using data for the unpolarized RCS cross section below threshold.
The experimental fits shown by black curves have been obtained within unsubtracted DRs~\cite{Lvov:1996rmi,OlmosdeLeon:2001zn}.  
Recently, covariant baryon chiral perturbation theory (BChPT)~\cite{Lensky:2015awa} and heavy baryon chiral perturbation theory (HBChPT)~\cite{McGovern:2012ew} have been developed as a convenient framework to analyze RCS data up to the $\Delta$-resonance region. The corresponding fits are shown by the brown disk and blue curve, respectively, and  yield values for the magnetic polarizabilities larger than the fits with DRs. The results from the analysis of HBChPT~\cite{McGovern:2012ew} was recently also included in the PDG average (violet disk), 
resulting in the following  values~\cite{Patrignani:2016xqp}:
\bea
\alpha_{E1}=(11.2\pm0.4)\times 10^{-4}\, \mathrm{fm}^3,\quad \beta_{M1}=(2.5\mp0.4)\times 10^{-4}\, \mathrm{fm}^3.\label{alpha-beta-PDG}
\eea
Finally we quote the results from a recent fit within subtracted DRs~\cite{Pasquini:2017ehj}, which uses  as input the MAID07 pion-photoproduction multipoles~\cite{Drechsel:2007if}, and the  values for the spin polarizabilities  extracted from double polarization RCS measurements at MAMI~\cite{Martel:2014pba}.
They have been obtained with and without the constraint of the Baldin sum rule, corresponding respectively to the closed and open red disks in \textbf{Figure \ref{fig1}}:
\bea
\alpha_{E1}&=&(11.91\pm0.22)\times 10^{-4}\, \mathrm{fm}^3,\,\, \beta_{M1}=(1.86\mp0.22)\times 10^{-4}\,\, \mathrm{fm}^3\, \mathrm{(w/} \, \mathrm{Baldin}\, \mathrm{SR)},\\
\alpha_{E1}&=&(11.63\pm0.38)\times 10^{-4}\, \mathrm{fm}^3,\,\, \beta_{M1}=(1.49\pm0.48)\times 10^{-4}\,\, \mathrm{fm}^3\, \mathrm{(w/o} \, \mathrm{Baldin}\,\mathrm{SR)}.
\eea
These results clearly show that the tension for $\beta_{M1}$ between the fits within effective field theories and DRs persists. One should also note that the various fits in \textbf{Figure \ref{fig1}} have been obtained
using different data sets. For example, Ref.~\cite{McGovern:2012ew} has defined an ``improved" data set, where a few data points from different experiments have been discarded, whereas the fit with subtracted DRs includes the ``full" data set consisting of all available data for the unpolarized RCS cross section  below  threshold,  i.e., the more recent data of Refs.~\cite{OlmosdeLeon:2001zn,Zieger:1992jq,Federspiel:1991yd,MacGibbon:1995in,Hallin:1993ft} as well as the older data listed in~\cite{Pasquini:2017ehj}.
Preliminary studies of the statistical consistency of the different data subsets show that the fit results may depend on the choice of the data set~\cite{Pasquini:2017ehj,Krupina:2017pgr}, and therefore the comparison between various fits is not conclusive at the moment. 
Future measurements planned at MAMI~\cite{Martel:2017pln} and at the High Intensity Gamma-Ray Source (HI$\gamma$S)~\cite{Weller:2009zza} hold the promise to clarify this situation. In particular, measurements are underway at MAMI~\cite{A2-Downie}  of both the unpolarized cross section and beam asymmetry, with the aim to extract  the proton scalar polarizabilities with unprecedented precision from a single experiment.
First results of the beam asymmetry with lower statistics provide a proof-of-principle that the scalar polarizabilities  can be accessed in this way~\cite{Sokhoyan:2016yrc}.

In contrast to the scalar polarizabilities, much less is known for the spin polarizabilities.
In addition to the results for the FSP from the GGT sum rule discussed in Section~\ref{FRCS}, the experimental value for the backward spin polarizability $\gamma_\pi$ has been obtained by an analysis with unsubtracted DRs of
backward angle Compton scattering.
The average value from
three measurements at MAMI (TAPS~\cite{OlmosdeLeon:2001zn}, LARA~\cite{Galler:2001ht,Wolf:2001ha}, and SENECA~\cite{Camen:2001st})
yields:
 \bea
 \gamma_\pi=(8.0\pm 1.8)\times 10^{-4}\,\mathrm{fm}^4.\label{gpi-exp}
 \eea
To obtain information on the individual spin polarizabilities, one has to resort to double polarization experiments.
A systematic study of the sensitivity to the individual polarizabilities of the unpolarized and double polarized Compton observables, with beam and target polarizations, has been performed using subtracted DRs~\cite{Pasquini:2007hf},  and has been used  to plan the  double-polarization experimental program at MAMI~\cite{A2-Hornidge}.
More recently, this analysis has been complemented by a study using  BChPT~\cite{Griesshammer:2017txw}.
In  \textbf {Table \ref{tab2}} we show the results  from the  fit, within subtracted DRs, of the recent MAMI measurements~\cite{Martel:2014pba} for the double polarization asymmetry using circularly polarized  photons and transversely polarized
proton target ($\Sigma_{2x}$) along with the data for the beam asymmetry  either from the LEGS experiment ($\Sigma_3^{{\rm LEGS}}$)~\cite{Blanpied:2001ae} or from the recent MAMI experiment ($\Sigma_3^{{\rm MAMI}}$)~\cite{Sokhoyan:2016yrc}.
These data have been analyzed to extract the spin polarizabilities $\gamma_{E1E1}$ and $\gamma_{M1M1}$, while the remaining leading static polarizabilities
were constrained within the uncertainties of  the available experimental results.
In particular,  the scalar polarizabilities were taken from  the 2012 PDG values~\cite{Beringer:1900zz}, i.e. $\alpha_{E1}=(12.16\pm0.58)\cdot10^{-4}$ fm$^3$  and $\beta_{M1}=(1.66\pm0.69)\cdot10^{-4}$ fm$^3$,  $\gamma_0$ from the GDH \& A2 value in \textbf {Table \ref{tab1}} and $\gamma_\pi$ from the experimental value~\ref{gpi-exp}.  
The data for $\Sigma_{2x}$ and $\Sigma_3^{{\rm LEGS}}$ have also been analyzed in~\cite{Martel:2014pba} within BChPT, giving values compatible, within uncertainties,  with the DR fit. This is a positive indication that the model dependence of the polarizability fitting is comparable to, or smaller than, the statistical errors of the data.
In \textbf {Table \ref{tab2}} we also report  the predictions of unsubtracted DRs, obtained by evaluating the non-Born amplitudes in Eq.~\ref{eq:unsub} at $\nu=t=0$ with the MAID07 input, and the calculations within HBChPT~\cite{McGovern:2012ew}, BChPT~\cite{Lensky:2015awa} and a chiral Lagrangian approach (L$_\chi$)~\cite{Gasparyan:2011yw}. The uncertainties in the fit values are still too large to discriminate between the various approaches.
Further analyses, with an unconstrained fit of all the six leading static polarizabilities, including the  MAMI measurements for the beam asymmetry~\cite{A2-Downie} and double-polarization asymmetry with circularly polarized  photons and longitudinally polarized target~\cite{A2-Hornidge}, hold the promise  to pin down the values for the individual spin polarizabilities with better precision.

\begin{table}[t]
\tabcolsep7.5pt
\caption{Results for the static spin polarizabilities (in units of $10^{-4}$ fm$^4$)  from the fit with subtracted DRs
to $\Sigma_{2x}$~\cite{Martel:2017pln,Martel:2014pba} along with either $\Sigma_3^{{\rm LEGS}}$~\cite{Blanpied:2001ae} (first column) or $\Sigma_3^{{\rm MAMI}}$ (second column), in comparison with predictions from  unsubtracted DRs with the  MAID07 input, 
HBChPT~\cite{McGovern:2012ew},  BChPT~\cite{Lensky:2015awa} and  a chiral Lagrangian (L$_\chi$)~\cite{Gasparyan:2011yw}.}
\label{tab2}
\begin{center}
\begin{tabular}{@{}l|c|c|c|c|c|c@{}}
\hline
 &$\Sigma_{2x}$ and $\Sigma_3^{{\rm LEGS}}$ &$\Sigma_{2x}$ and $\Sigma_3^{{\rm MAMI}}$&DRs &HBChPT&BChPT&L$_\chi$\\
\hline
$\gamma_{E1E1}$  & $-3.5\pm 1.2$     &  $-5.0\pm1.5$   & $-4.5$  & $-1.1\pm1.8$&$-3.3 \pm 0.8$&$-3.7$ \\
$\gamma_{M1M1}$ & $3.16\pm 0.85$  &  $3.13\pm0.88$ & 3.0&$2.2\pm0.7$$^{\rm a}$ & $2.9 \pm 1.5$&2.5\\
$\gamma_{E1M2}$ & $-0.7\pm 1.2$     & $1.7\pm1.7$     &  $-0.08$&$-0.4\pm0.4$ & $0.2 \pm 0.2$&1.2\\
$\gamma_{M1E2}$ & $1.99\pm0.29$   & $1.26\pm0.43$ &2.3 &$1.9\pm0.4$ &   $1.1 \pm 0.3$ &1.2\\
\hline
\end{tabular}
\end{center}
\begin{tabnote}
$^{\rm a}$ An additional error of $\pm 0.5$ comes from the fit of the $\gamma N\Delta$ coupling constant  to RCS data~\cite{McGovern:2012ew}.
\end{tabnote}
\end{table}

\subsection{Dynamical polarizabilities}
\label{RCS-dyn-pol}

\begin{figure}[t]
\includegraphics[width=0.625\textwidth]{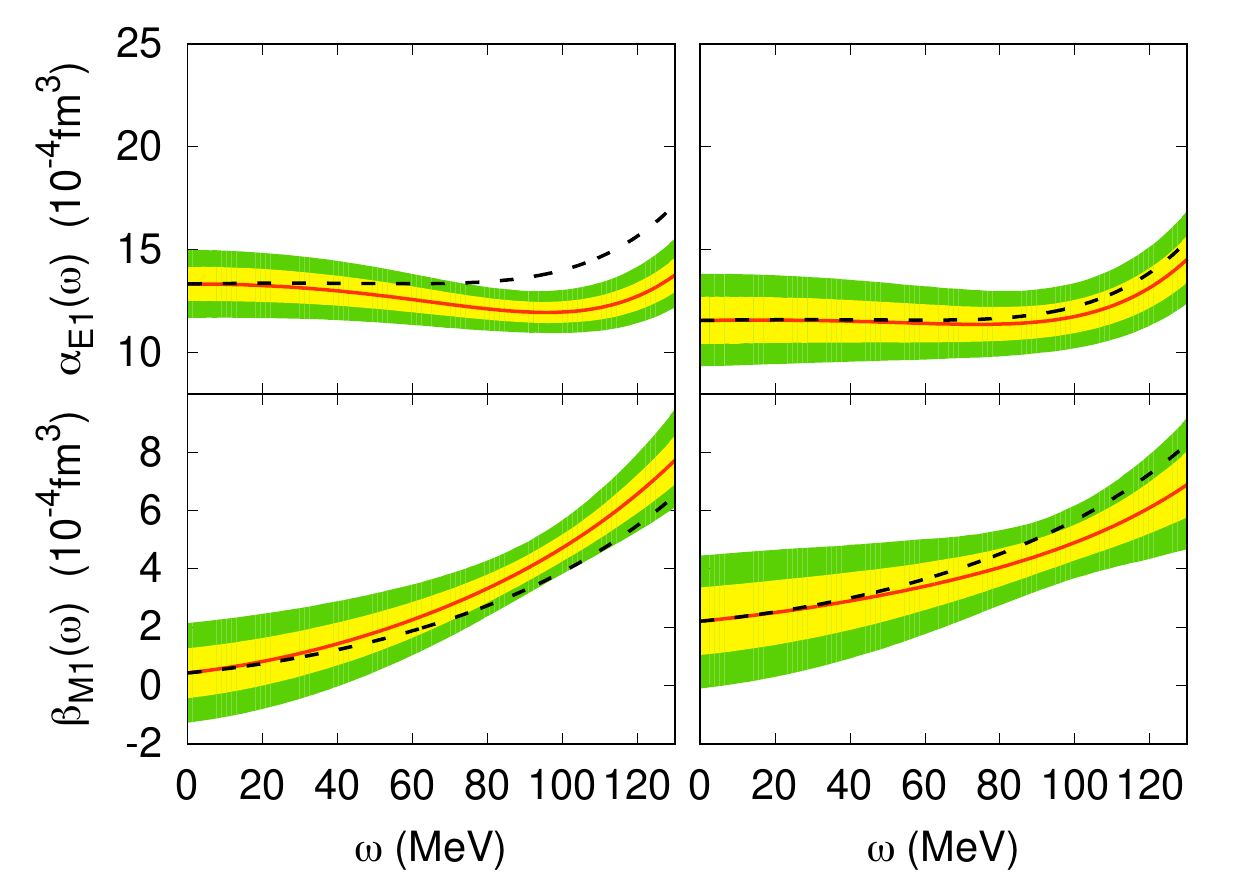}
\caption{Results from the fit   of the DDPs (red solid line) as function of the cm photon energy $\omega$, using the full data set (left panels) and the TAPS subset (right panels): 
$\alpha_{E1}(\omega)$   on the top and $\beta_{M1}(\omega)$ on the bottom, with  the $68\%$ 
  (yellow)
  and $95\%$  (green) C.L. areas. The dashed lines are predictions from subtracted DRs, with the  values at $\omega=0$ fixed to the fit results.
Figure  from Ref.~\cite{Pasquini:2017ehj}.}
\label{fig3}
\end{figure}

Dynamical polarizabilities combine the concepts of multipole expansion of the scattering amplitudes and nucleon polarizabilities and provide
 a better filter for the mechanisms governing the nucleon response in Compton scattering. They are functions of the excitation energy and encode the dispersive effects
of $\pi N$, $N^*$  and other higher intermediate states~\cite{Babusci:1998ww,Griesshammer:2001uw,Hildebrandt:2003fm}. 
  The information encoded in the dynamical nucleon polarizabilities has been pointed out in different theoretical
calculations, using DRs or effective field
theories~\cite{Lensky:2009uv,Griesshammer:2001uw,Hildebrandt:2003fm}. 
  However, extracting these polarizabilities from RCS data is very challenging,  because of the very low sensitivity of the  RCS data to the higher-order dispersive coefficients and the strong correlations between the fit parameters.
  Work in this direction has been presented recently in~\cite{Pasquini:2017ehj}, where first information on the scalar dynamical dipole polarizabilities (DDPs) has been extracted from RCS data below threshold. 
The theoretical framework for such analysis relies on the multipole expansion of the scattering amplitude, the LEX of the DDPs, and subtracted DRs for the calculation of the higher-order multipole amplitudes. The  statistical analysis was performed using a new method  based on the bootstrap
technique, that turned out to be crucial to deal with
problems inherent to both the low sensitivity of the RCS cross section to the energy dependence of the DDPs and to the limited  accuracy of the available data sets.
  The results of such analysis are shown in \textbf{Figure \ref{fig3}} (red solid curve), with $68\%$ 
  (yellow)  and $95\%$ (green) confidence level (C.L.)  uncertainty bands. They have been obtained using  two different data sets, i.e., 
  the full data set  and  the  data set given by the TAPS experiment alone~\cite{OlmosdeLeon:2001zn}, which is, by far, the most comprehensive available subset.
The fit results are compared with the subtracted DR predictions~\cite{Hildebrandt:2003fm} (dashed curves) using the MAID07 input~\cite{Drechsel:2007if}. 
At zero energy, one recovers the value of the static electric and magnetic dipole polarizabilities.
At  energy $\omega\lesssim 60$  MeV the DR results are within the $68\%$ confidence area of the fit results for all the DDPs. At higher energies, 
 the DR predictions for $\beta_{M1}(\omega)$ remain within the 
$95\%$ C.L. region, while for $\alpha_{E1}(\omega)$ we observe deviations from the fit results in the case of the full data set and a very good agreement, 
within the $68\%$ confidence area, in the case of the TAPS data set. This different behavior can be a hint of inconsistencies between the two data sets. The larger error bands in the case of $\beta_{M1}(\omega)$ also reflect the lower sensitivity of the unpolarized RCS data to the magnetic as compared to the electric polarizability.
The high-precision measurements  planned
at MAMI below pion-production threshold~\cite{A2-Downie} will definitely help to disentangle with better accuracy the effects of the individual leading-order static and higher-order dispersive polarizabilities.

\section{Virtual Compton scattering}
\label{VCS}

Virtual Compton scattering (VCS) is formally obtained from RCS  by replacing the incident real
photon with a virtual photon $\gamma^*$, and 
can be accessed experimentally as a subprocess of the reaction $e(k)+N(p)\rightarrow e(k')+N(p')+\gamma(q')$, where the real final photon can be emitted by either the electron or the nucleon.
The first process corresponds to the Bethe-Heitler (BH) contribution, which is well known and entirely calculable from QED with the nucleon electromagnetic  form factors as input.
The second one contains, in the one-photon exchange approximation, the VCS  subprocess.
The VCS contribution can be further decomposed in a Born term, 
where the intermediate state is a nucleon as defined in~\cite{Guichon:1995pu}, and a non-Born term, which contains all nucleon excitations and meson-loop contributions.
At low energy $q'$ of the emitted photons, one can use the LET for VCS~\cite{Guichon:1995pu,Scherer:1996ux,Low:1954kd}, which states that
the non-Born term starts at order $q'$, whereas the  Born term enters at  $1/q'$.
If we parametrize the non-Born term   with a multipole expansion in the cm  of the $\gamma^* N$ system, 
the leading contribution in $q'=|\boldsymbol{q}'|_{{\rm cm}}$ can be expressed in terms of generalized polarizabilities (GPs) with the
 multipolarities of the emitted photon corresponding to electric and magnetic dipole radiation.
 For a dipole transition in the final state and arbitrary three-momentum $q=|\boldsymbol{q}|_{{\rm cm}}$ of the virtual photon, angular momentum and parity conservation lead
to ten GPs~\cite{Guichon:1995pu}, depending  on the virtuality $Q^2$ of the virtual photon.
By further imposing nucleon crossing and  charge
conjugation symmetry, the number of independent GPs reduces to six~\cite{Drechsel:1997xv}. They correspond to two scalar GPs, 
$\alpha_{E1}(Q^2)$ and $\beta_{M1}(Q^2)$, reducing to the RCS static scalar polarizabilities in the limit of $Q^2\rightarrow 0$, and four spin-dependent GPs, denoted as~\cite{Guichon:1998xv}: 
\begin{equation}
 P^{(L1,L1)1}(Q^2),\quad P^{(M1,M1)1}(Q^2),\quad
P^{(M1,L2)1}(Q^2),\quad P^{(L1,M2)1}(Q^2). 
\label{spinGPs}
\end{equation}
In this notation, $P^{(\rho'\ell',\rho\ell)1}$ corresponds with 
a multipole amplitude where $\rho =L, M$ denotes whether the photon is of longitudinal or magnetic type and $\ell$ denotes the angular momentum
(respectively, $\rho'\ell'$ or $\rho\ell$ for the final or initial photon); the index $1$ at the end indicates that the transition involves a 
nucleon spin flip.
In the limit $Q^2\rightarrow 0$, the first two spin GPs in Eq.~\ref{spinGPs} vanish, whereas the latter two reduce to 
the RCS spin polarizabilities as~\cite{Drechsel:1998zm}:
\begin{equation}
\gamma_{E1 M2} = - \alpha_\mathrm{em} \frac{3}{\sqrt{2}} \, P^{(L1, M2)1}(0), \quad 
\gamma_{M1 E2} = - \alpha_\mathrm{em} \frac{3 \sqrt{3}}{2 \sqrt{2}} \, P^{(M1, L2)1}(0), 
\end{equation}
where $\alpha_\mathrm{em}\equiv e^2/4\pi\simeq 1/137$ denotes the fine-structure constant.

According to the LET, 
the LEX of the VCS observables  provides a method to analyze VCS experiments below pion-production threshold
 in terms of structure functions which contain information on  GPs~\cite{Guichon:1995pu,Guichon:1998xv}. 
However, the  sensitivity of the VCS cross section to the GPs is enhanced in the region between pion-production threshold and the $\Delta$-resonance region.
The LEX does not hold in this regime, but the dispersive
approach is expected to give a reasonable framework to extract the GPs.
To set up the DR formalism, we can parametrize the non-Born contribution to the VCS scattering amplitude in terms of twelve independent amplitudes $F_i(Q^2,\nu,t), \, i=1,\dots, 12$, free of kinematical singularities and constraints and even in $\nu$~\cite{Pasquini:2001yy}. Furthermore, the GPs are expressed in terms of the
non-Born part $F^{NB}_i$  at the point $t=-Q^2$ and $\nu=0$.
Assuming  an appropriate analytic and high-energy
behavior, these amplitudes fulfil unsubtracted
DRs in the variable $\nu$ at fixed $t$ and fixed $Q^2$:
\bea
\mathrm{Re}\, F^{NB}_{i}(Q^2,\nu,t)=F_{i}^{pole}(Q^2,\nu,t)-F^{B}_{i}(Q^2,\nu,t)+ \frac{2}{\pi}\mathcal{P}\int_{\nu_{thr}}^{+\infty}{\rm d}\nu'\frac{\nu'\mathrm{Im}\, F_i(Q^2,\nu',t)}{\nu'^2-\nu^2},
\label{DR-VCS}
\eea
where $F_{i}^B$ is the Born contribution as defined in~\cite{Guichon:1995pu,Guichon:1998xv}, whereas
$F_{i}^{pole}$ denote the nucleon pole contributions. Furthermore, $\mathrm{Im} \,F_i$ are the
discontinuities across the $s$-channel cuts, starting at the
pion production threshold $\nu_{thr}=m_\pi+(m_{\pi}^2+t/2+Q^2/2)/(2M)$.

The validity of the  unsubtracted DRs in Eq.~\ref{DR-VCS} relies on the assumption that at high energies ($\nu\rightarrow\infty$,
 fixed $t$ and fixed $Q^2$) the amplitudes drop fast enough such that the integrals converge.
The high-energy behavior of the amplitudes $F_i$
was investigated in~\cite{Pasquini:2000pk,Pasquini:2001yy}, with the finding that the integrals diverge for 
$F_1$ and $F_5$. 
As long as we are interested in the
energy region up to the $\Delta$-resonance, we may saturate the
$s$-channel dispersion integral by the $\pi N$ contribution, setting the upper limit of integration to $\nu_{max}=1.5 $ GeV.
The remainder can be estimated by energy-independent functions, which parametrize the
asymptotic contribution due to $t$-channel poles, as well as the residual dispersive contributions
beyond the value $\nu_{max} = 1.5$ GeV.
The asymptotic contribution to $F_5$ is saturated by
the $\pi^0$ pole~\cite{Pasquini:2001yy}.
The asymptotic contribution to  $F_1$ can be described phenomenologically  as the  exchange of an effective $\sigma$ meson, in the same spirit as for unsubtracted DRs in the RCS case.  The $Q^2$ dependence of this term is unknown. It can be parametrized in terms of a function directly related to the magnetic dipole GP $\beta_{M1}(Q^2)$ and 
fitted to VCS observables.
Furthermore, it was found that the unsubtracted DR for the amplitude $F_2$ is not so well saturated by $\pi N$ intermediate states only. The additional $s$-channel contributions  beyond the $\pi N$ states can effectively be accounted for with an energy-independent function, at fixed $Q^2$ and $t=-Q^2$.
This amounts to introducing an additional fit function, which is directly related to the electric dipole GP $\alpha_{E1} (Q^2)$.
 In order to provide predictions for VCS observables, it is convenient to adopt
the following  parametrizations for the fit functions:
\bea
\alpha_{E1}(Q^2)-\alpha_{E1}^{\pi N}(Q^2)=\frac{\alpha_{E1}^{exp}-\alpha_{E1}^{\pi N}}{(1+Q^2/\Lambda_\alpha^2)^2},\quad
\beta_{M1}(Q^2)-\beta_{M1}^{\pi N}(Q^2)=\frac{\beta_{M1}^{exp}-\beta_{M1}^{\pi N}}{(1+Q^2/\Lambda_\beta^2)^2},
\label{GPs-parametrization}
\eea
where $\alpha_{E1}$ and  $\beta_{M1}$ are the RCS polarizabilities, with superscripts $exp$ and $\pi N$ indicating, respectively, the experimental value and the $\pi N$ contribution evaluated from unsubtracted DRs.
In Eq.~\ref{GPs-parametrization},  the mass scale parameters $\Lambda_\alpha$ and $\Lambda_\beta$ are  free parameters, not necessarily constant with $Q^2$, which can be adjusted by a fit to the experimental cross sections.

\begin{figure}[t]
\centerline{%
\includegraphics[width = .5 \textwidth]{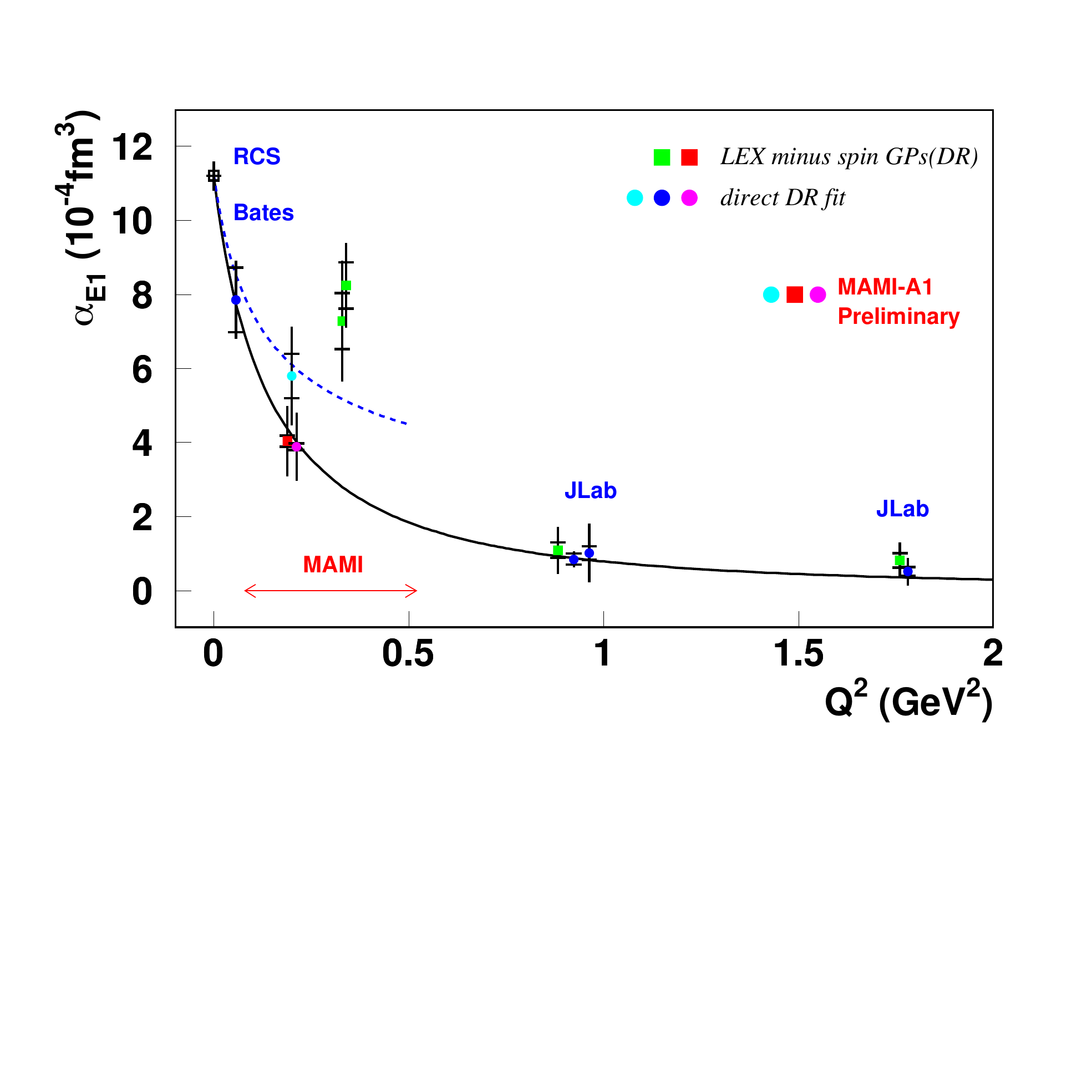}%
\hspace{-6 cm}
\includegraphics[width = .5 \textwidth]{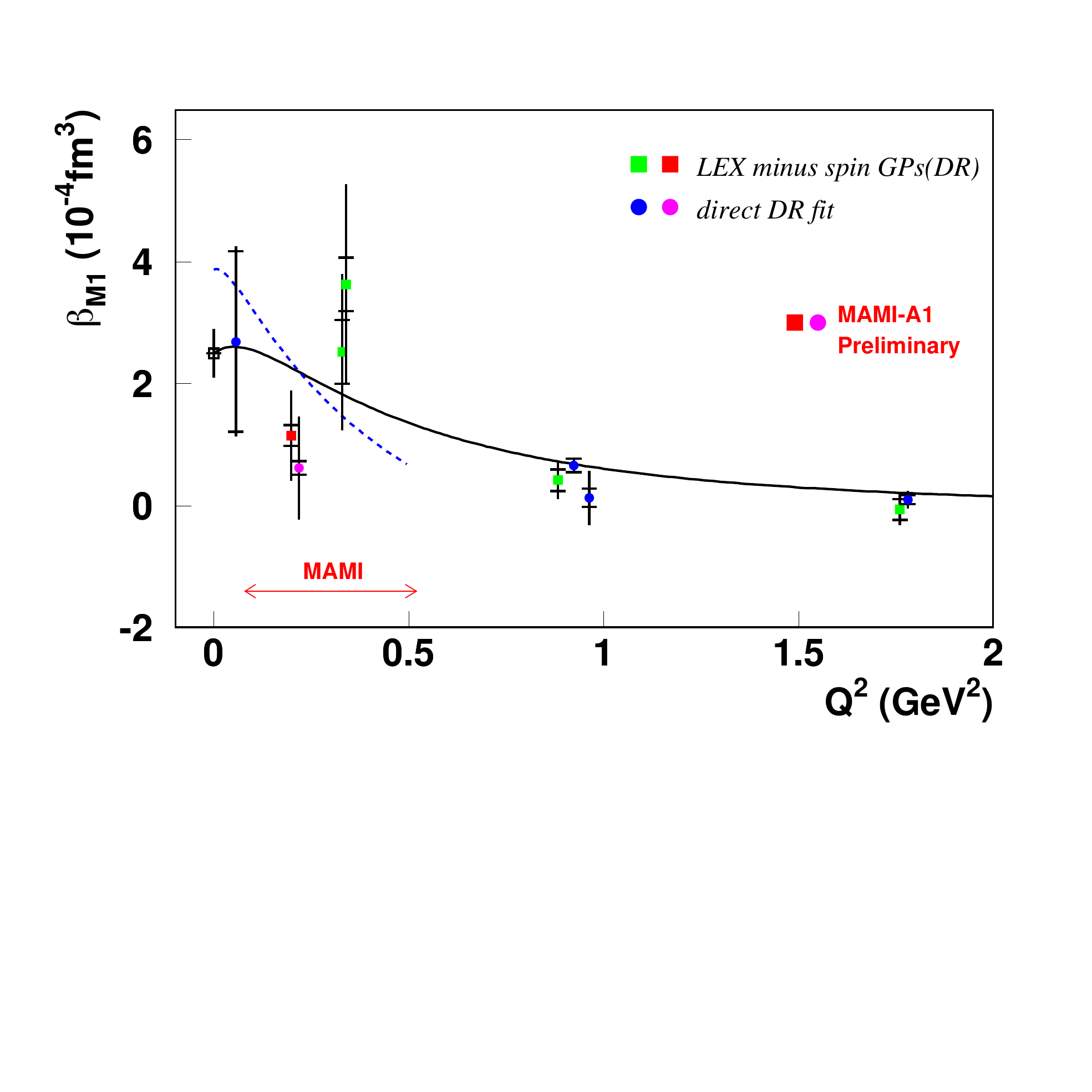}
\vspace{-2cm}
}
\caption{Compilation of world data for $\alpha_{E1}(Q^2)$ (left) and $\beta_{M1}(Q^2)$ (right) at different $Q^2$ values in GeV$^2$:
$Q^2 =0$ from the PDG values, Eq.~\ref{alpha-beta-PDG};   $Q^2 = 0.06$ from MIT-Bates~\cite{Bourgeois:2006js,Bourgeois:2011zz}; $Q^2 = 0.33$ from MAMI~\cite{Roche:2000ng,Janssens:2008qe};
 $Q^2 = 0.2$ from the preliminary analysis of  MAMI ~\cite{A1-VCS,Correa,Blomberg}; $Q^2 = 0.92, 1.76$ from  JLab~\cite{Laveissiere:2004nf,Fonvieille:2012cd}.
Solid curves: DR predictions, with mass scales $\Lambda_\alpha = 0.73$ GeV and $\Lambda_\beta=0.63$ GeV. Dashed curves: BChPT predictions~\cite{Lensky:2016nui}.
Plot courtesy of H. Fonvieille.
}
\label{fig4}
\end{figure}

A series of VCS measurements at MAMI~\cite{Roche:2000ng,Bensafa:2006wr,Janssens:2008qe,Doria:2015dyx,A1-VCS,Correa,Blomberg},
JLab ~\cite{Laveissiere:2004nf,Fonvieille:2012cd}, and Bates~\cite{Bourgeois:2011zz,Bourgeois:2006js}
have provided a first experimental exploration of the proton's electric and magnetic GPs.
These experiments involve measurements below and above pion threshold, and results have
been extracted using  both the LEX and  DR approaches. 
A fundamental difference between the two analysis methods is that the DR formalism allows for a direct extraction of the scalar GPs, by fitting the parameters $\Lambda_\alpha$ and $\Lambda_\beta$ to the data, whereas the LEX analysis gives access  to structure functions depending linearly 
 on both scalar and spin GPs~\cite{Guichon:1995pu}. In order to disentangle the scalar GPs, the contribution from the spin GPs to the structure functions has to be subtracted using a model.
One usually uses the DR model, introducing some model dependence which is presently not accounted for in the error bars. \textbf{Figure \ref{fig4}} displays the results for the electric GP, $\alpha_{E1}(Q^2)$, and the magnetic GP, $\beta_{M1}(Q^2)$, from the world VCS measurements, showing a nice consistency between the LEX and DR extractions.
The solid curves correspond to the DR predictions, obtained with the PDG values of Eq.~\ref{alpha-beta-PDG}  for the static polarizabilities at $Q^2=0$ and the mass scales 
$\Lambda_\alpha = 0.73$ GeV and $\Lambda_\beta=0.63$ GeV. The dashed curves are the predictions from BChPT~\cite{Lensky:2016nui}, which are plotted in the low-$Q^2$ range of applicability of the theory and without the theory-uncertainty band.

One notices from \textbf{Figure \ref{fig4}} that the electric GP,  which is dominated by the asymptotic contribution, 
cannot be described by a single dipole form over the full $Q^2$ range.
In particular, the data situation  near $Q^2 = 0.3$ GeV$^2$ is currently not
understood, since all  the models, such as  chiral effective field theories~\cite{Hemmert:1996gr,Hemmert:1997at,Hemmert:1999pz,Kao:2002cn,Kao:2004us,Lensky:2016nui}, the linear-$\sigma$ model~\cite{Metz:1996fn,Metz:1997fr}, non-relativistic~\cite{Liu:1996xd,Pasquini:2000ue} and relativistic~\cite{Pasquini:1997by} constituent quark models,
predict a smooth fall-off with $Q^2$. 
The magnetic GP results from a large dispersive $\pi N$ (paramagnetic) contribution, 
dominated by $\Delta(1232)$ resonance, and a large asymptotic (diamagnetic) contribution
with opposite sign, leading to a relatively small net result with a relatively flat behavior at low $Q^2$.
We also note the difference at low $Q^2$ between the DR and BChPT predictions, which are not resolved by the existing experimental data.
More high-precision measurements are needed, and the new experimental data from MAMI~\cite{A1-VCS} at  $Q^2=0.1$ and $0.45$ GeV$^2$ together  with the upcoming measurements at JLab~\cite{JLab_C12-15-001} in the $Q^2$ range of 0.3-0.75 GeV$^2$ should mark a step forward in our understanding of the underlying mechanisms which govern the structure of the GPs at low and intermediate $Q^2$.

\begin{figure}[h]
\centerline{%
\includegraphics[width = .49 \textwidth]{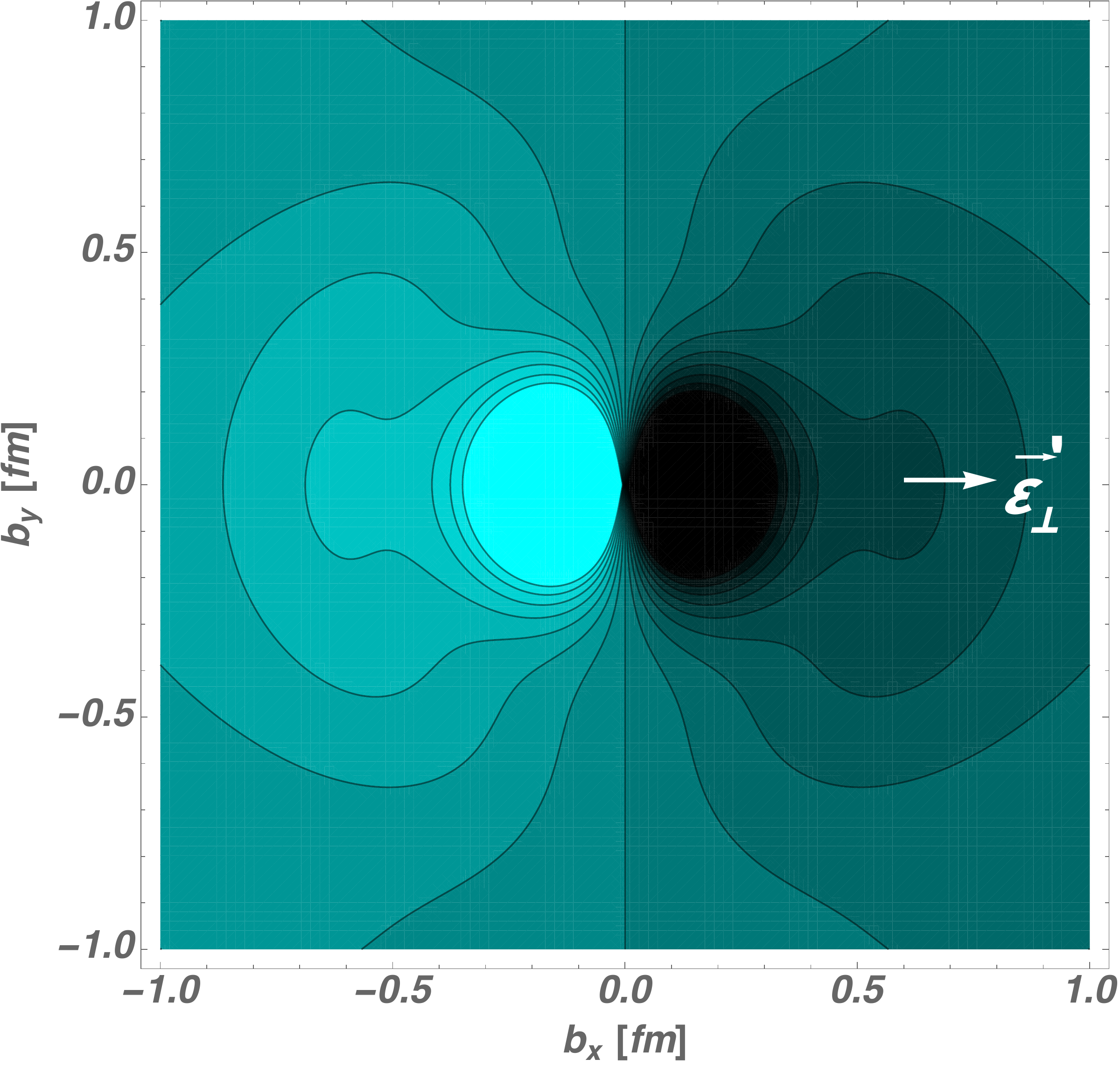}%
\hspace{-6.5cm}
\includegraphics[width = .49 \textwidth]{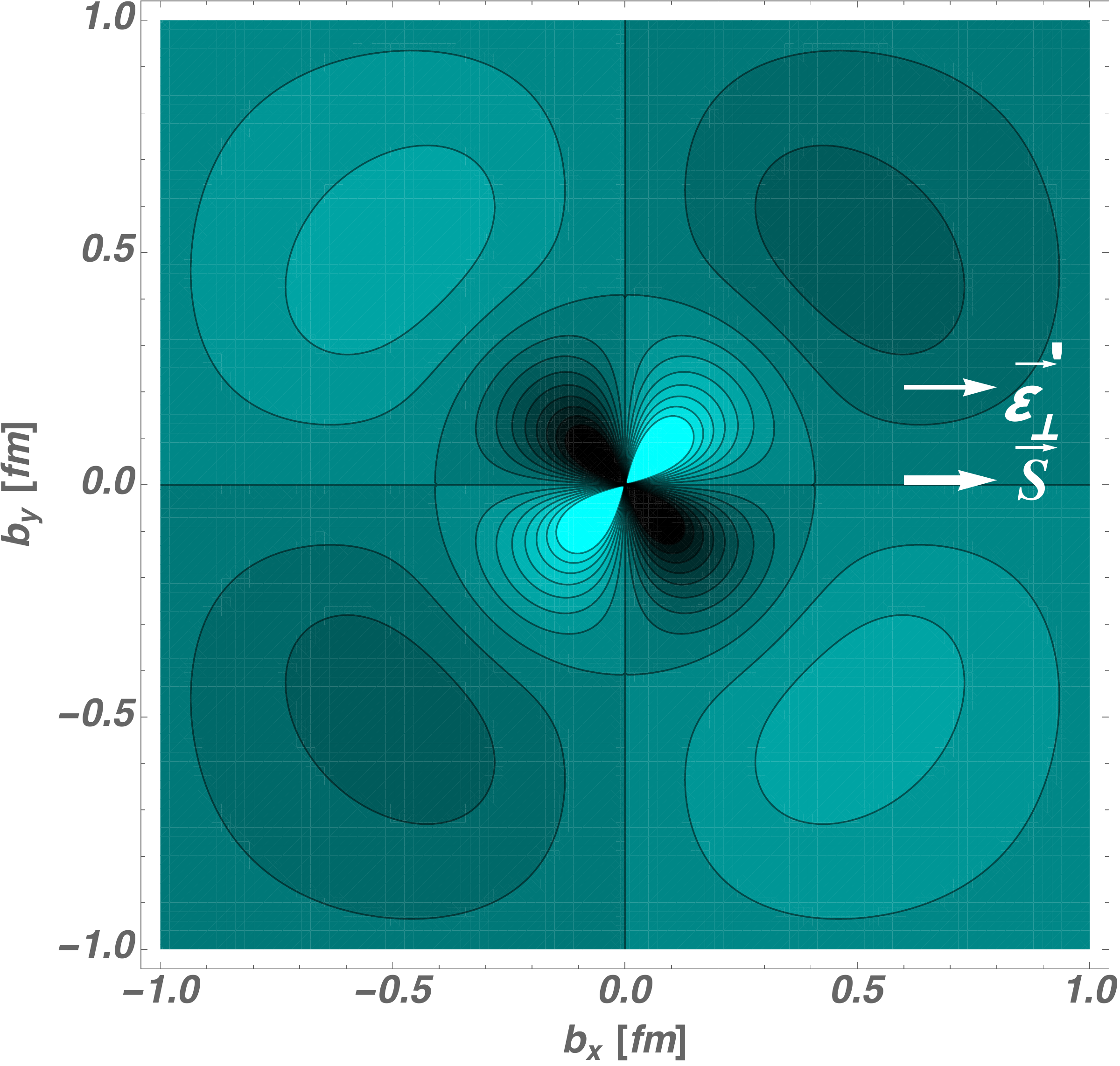}%
}%
\vspace{0.25 truecm}
\caption{Induced polarization density  in a proton of 
definite light-cone helicity (left), and with spin $\boldsymbol{S}$ along the $x$-axis (right), when submitted to an electromagnetic field 
 with photon polarization along the $x$-axis, 
as indicated. Light (dark) regions correspond to the
largest (smallest) values.
}\label{fig5}
\end{figure}
A precise knowledge of the $Q^2$ dependence of the GPs is  crucial  to obtain, through a Fourier transform, a spatial representation 
 of the deformation of the charge and magnetization distributions  of the nucleon under the influence of an external static electromagnetic field~\cite{Lvov:2001zdg,Gorchtein:2009qq}.
A proper spatial-density interpretation can be formulated  by considering the nucleon in a light-front frame~\cite{Gorchtein:2009qq}. In this frame, the two transverse components  $\boldsymbol{q}_\perp$ of the virtual
photon momentum,  with $Q^2=\boldsymbol{q}_\perp^{2}$, are the conjugate variables to the transverse position coordinates $\boldsymbol{b}_\perp$ of the quarks in a nucleon.
Furthermore, the polarization vector $\boldsymbol{\epsilon}'_\perp$ of the outgoing photon is associated with an external quasi-static electromagnetic field, polarizing the charge and magnetization distributions.
Such polarization densities are described by different scalar and spin GPs, depending on the spin state of the nucleon.
 In~\textbf{Figure \ref{fig5}} we show the polarization densities in transverse-position space for a proton in a definite light-front helicity state (left panel) and for a proton 
in an eigenstate of the transverse-spin $\boldsymbol{S}$ (right panel), as calculated using the DR results of the GPs shown in  \textbf{Figure \ref{fig4}}.
In the former case, the polarization density displays a  dipole pattern dominated by the scalar GPs, with the spatial extension at the nucleon periphery  strongly depending on the mass scale $\Lambda_{\alpha}$.
In the second case, on top of a weak dipole deformation, we observe a quadrupole pattern
 with pronounced strength around 0.5 fm due to the electric GP.

\section{Forward double virtual Compton scattering}
\label{VVCS}

In this section, we review the application of DRs to the {\it forward} double VCS (VVCS) process:  
\bea
\gamma^\ast (q) + N(p) \to \gamma^\ast(q) + N(p),
\label{doublevcs}
\eea
where both photons have the same finite space-like virtuality $q^2 = - Q^2 \leq 0$. We discuss the resulting LEX for the VVCS amplitudes and sum rules in terms of generalized, i.e. $Q^2$ dependent, polarizabilities. 
For the polarized VVCS amplitude, we discuss two model-independent relations at low $Q^2$, connecting moments of spin structure functions to polarizabilities accessible in RCS and VCS, as discussed in Sections~\ref{RCS} and ~\ref{VCS} respectively. 
For the unpolarized VVCS amplitude involving transverse virtual photons, one subtraction function is required. We discuss the information on the $Q^2$ dependence of this subtraction function in terms of polarizabilities and chiral effective field theories.

The forward VVCS tensor $M^{\mu \nu}$, with $\mu$ $(\nu)$ denoting the four-vector index of initial (final) photon, is described by four invariant amplitudes, denoted by $T_1, T_2, S_1, S_2$, which are functions 
of $Q^2$ and $\nu \equiv p \cdot q / M$, as~\cite{Lensky:2017bwi}: 
\bea
\label{vvcs}
\alpha_{em} \, M^{\mu \nu} &=&
\left( g^{\mu\nu} - \frac{q^{\mu}q^{\nu}}{q^2}\right)
T_1(\nu, Q^2) 
- \frac{1}{M^2} \left(p^{\mu}-\frac{p\cdot
q}{q^2}\,q^{\mu}\right) \left(p^{\nu}-\frac{p\cdot
q}{q^2}\, q^{\nu} \right) T_2 (\nu, Q^2) \nn \\
&-& \frac{i}{M} \epsilon^{\nu\mu\alpha\beta}\,q_{\alpha}
s_{\beta}\, S_1(\nu, Q^2) 
- \frac{i}{M^3} \epsilon^{\nu\mu\alpha\beta}\,q_{\alpha}
(p\cdot q\ s_{\beta}-s\cdot q\ p_{\beta})\, S_2 (\nu, Q^2)  ,
\eea
where $\epsilon_{0123} = +1$, $s^\alpha$ is the nucleon covariant
spin vector satisfying $s \cdot p$ = 0, $s^2 = -1$. 
Note that this definition implies that at the real photon point the amplitudes $T_1$ and $S_1$ are related 
to the amplitudes $f$ and $g$ of 
Eq.~\ref{frcs}, describing the forward RCS,  as: $T_1(\nu, 0) = f(\nu)$ and $S_1(\nu, 0) = (M/\nu) g(\nu)$. 

The optical theorem yields  the following relations for the  imaginary parts of the four
amplitudes appearing in Eq.~\ref{vvcs}:
\bea
\label{optical}
{\rm{Im}}\ T_1(\nu,\,Q^2) = \frac{e^2}{4 M} F_1(x,\,Q^2) \,&,& \quad
{\rm{Im}}\ T_2(\nu,\,Q^2)  = \frac{e^2}{4 \nu}  F_2(x,\,Q^2) \, , \nn \\
{\rm{Im}}\ S_1(\nu,\,Q^2)  = \frac{e^2}{4 \nu}  g_1(x,\,Q^2) \, &,& \quad
{\rm{Im}}\ S_2(\nu,\,Q^2)  = \frac{e^2}{4} \frac{M}{\nu^2} g_2(x,\,Q^2) \, ,
\eea
where $x \equiv Q^2 / (2 M \nu)$ is the  Bjorken variable, and 
$F_1, F_2, g_1, g_2$ are the conventionally defined structure functions which parametrize 
inclusive electron-nucleon scattering. 
The imaginary parts of the forward scattering amplitudes, Eqs.~\ref{optical},
get contributions from both elastic scattering at $\nu = \nu_B  \equiv Q^2/(2M)$ or equivalently $x=1$, as well as 
from inelastic processes above pion threshold, corresponding with $\nu > \nu_{thr} \equiv m_\pi + (Q^2 + m_\pi^2)/(2 M)$ 
or equivalently $x < x_{thr} \equiv \nu_B /  \nu_{thr}$. 
The elastic contributions are obtained as pole parts of the direct and crossed nucleon Born diagrams. 
The latter are conventionally separated off the Compton scattering tensor in order to define 
structure dependent constants, such as polarizabilities. The Born terms are given by~\cite{Drechsel:2002ar}:
\bea
T_1^B  & = &
 -  \frac{\alpha_\mathrm{em}}{M}
(F_D^2 + \frac{\nu_B^2}{\nu^2-\nu_B^2+i \varepsilon} \,G_M^2) \,, \quad
T_2^B  =
- \frac{\alpha_\mathrm{em}}{M} \,\frac{Q^2}{\nu^2-\nu_B^2+i \varepsilon} \,
(F_D^2 + \tau \, F_P^2 ) \, ,\nn \\
S_1^B & = &
 -  \frac{\alpha_\mathrm{em}}{2M}
(F_P^2 + \frac{Q^2}{\nu^2-\nu_B^2+i \varepsilon}\,F_D G_M ) \,, \quad
S_2^B  =
\frac{\alpha_\mathrm{em}}{2}\frac{\nu}{\nu^2-\nu_B^2+i \varepsilon}
\,F_P G_M \, ,
\label{born}
\eea
where $\tau \equiv Q^2/ 4M^2$, and  $F_D$ and $F_P$ are the Dirac and Pauli form factors (FFs) of the nucleon $N$,
normalized to $F_D(0)=e_N$ and $F_P(0)=\kappa_N$. 
Furthermore, the magnetic FF combination is given by $G_M\,(Q^2) = F_D\,(Q^2)+F_P(Q^2)$.  
From the Born contributions of Eq.~\ref{born} one can directly read off the nucleon pole contributions.  

We next consider the analyticity in  $\nu$,  for fixed $Q^2$, of the  
VVCS amplitudes.  
We can distinguish two cases depending on the symmetry in the $s - u$ crossing variable $\nu$: 
$T_1, T_2$, and $S_1$ are even functions of $\nu$ whereas $S_2$ is 
an odd function of $\nu$. We discuss DRs for the non-pole  parts of the amplitudes,  
 i.e., when subtracting the 
well known pole contributions from the full amplitudes. 
In particular, we will consider
 unsubtracted DRs for the spin dependent amplitudes $S_1$ and $S_2$, and will then discuss the 
spin independent amplitudes, of which $T_1$ will require one subtraction. 
We will concentrate within the scope of 
this review on the model independent results at low $Q^2$. For a discussion of sum rules and of nucleon spin structure at larger $Q^2$, we refer the reader e.g. to the reviews of~\cite{Drechsel:2002ar,Kuhn:2008sy,Chen:2010qc}. 

\subsection{Spin dependent VVCS sum rules}

An unsubtracted DR for the non-pole ($np$) part of the amplitude $S_1$ is given by: 
\bea
\label{eq:S1dr}
{\rm{Re}} \,S_1^{np}(\nu,\,Q^2) \,=\,
2 \alpha_\mathrm{em} \,{\mathcal{P}}\,\int_{\nu_{thr}}^{\infty} \, d\nu' \, 
\frac{1}{\nu^{\prime \, 2} - \nu^2} \, \, g_1(x^\prime,\,Q^2),
\eea
with $x^\prime \equiv \nu_B /  \nu^\prime$. 
For a fixed finite value of $Q^2$, the LEX in $\nu$ for $S_1^{np}$ (and analogously for $S_2^{np}$) can be expressed through the 
inelastic odd moments of the structure functions $g_1$ ($g_2$), defined as:
\bea
\Gamma_{1, 2}^{(n) \, inel}(Q^2) &\equiv&  \int_{0}^{x_{thr}}\,dx \,  x^{(n-1)} \, g_{1, 2}\,(x,\,Q^2), \quad n = 1, 3, 5,...
\label{g2thirdmom} 
\eea
For $S_1^{np}$, the LEX takes the form~\cite{Ji:1999mr,Drechsel:2002ar}:
\bea
\label{S1lexnu}
S_1^{np}(\nu,\,Q^2) =
\frac{2 \, \alpha_\mathrm{em}}{M } \,  I_1(Q^2)
+ \left[ \frac{2 \, \alpha_\mathrm{em}}{\nu_B^3}  \Gamma_1^{(3) \, inel}(Q^2) \right] \, \nu^2  \,+\, {\mathcal{O}}(\nu^4),
\eea
where we have re-expressed the lowest moment as:
\bea
\label{I1sr}
I_1(Q^2) &\; \equiv \;& \frac{M}{\nu_B} \, \Gamma_1^{(1) \, inel}(Q^2) ,
\eea
which yields the GDH sum rule value at $Q^2 = 0$: $I_1(0) = - \kappa_N^2 / 4$. 
Furthermore at $Q^2 = 0$,  the term of ${\mathcal{O}}(\nu^2)$ in Eq.~(\ref{S1lexnu}) yields the FSP as: $\left[ (2  \alpha_\mathrm{em} / \nu_B^3)  \Gamma_1^{(3) \, inel}(Q^2) \right]_{Q^2 = 0} = M \gamma_0$. 
One can derive further sum rules by Taylor expanding Eq.~\ref{S1lexnu} for $S_1^{np}$ in $Q^2$ at $\nu = 0$. In this way, the following extension of the GDH sum rule to finite $Q^2$ was obtained in~\cite{Pascalutsa:2014zna, Lensky:2017dlc} 
for the slope at $Q^2 = 0$ of the moment $I_1(Q^2)$:
\bea
I_1^\prime(0) \equiv  \frac{d I_1}{d Q^2} \bigg|_{Q^2 = 0}= \frac{ \kappa_N^2}{12} \langle r_2^2 \rangle 
+ \frac{M^2}{2} \left\{ \frac{\gamma_{E1 M2}}{\alpha_\mathrm{em}}  
- 3 M \left[ P^{\prime  (M1, M1)1}(0) +  P^{\prime  (L1, L1)1} (0) \right]  \right\}. 
\label{qsqrgdhsr}
\eea
All quantities entering  Eq.~\ref{qsqrgdhsr} are observable quantities:  
the {\it lhs} is obtained from the first moment of the spin structure function $g_1$, 
whereas the {\it rhs} involves the squared Pauli radius $\langle r_2^2 \rangle $ 
as well as spin polarizabilities measured through the RCS and VCS processes, 
and introduced in Sections~\ref{RCS} and \ref{VCS}.

For the second spin-dependent VVCS amplitude $S_2$, which is odd in $\nu$, an 
unsubtracted DR takes the form:
\bea
\label{S2dr}
{\rm{Re}}\,   S_2(\nu,\,Q^2) 
\,=\,  2 \alpha_\mathrm{em} M \nu \,{\mathcal{P}}\,
\int_{0}^{\infty} d\nu^\prime  \frac{1}{\nu^{\prime \, 2} -\nu^2} \, \frac{1}{\nu^{\prime \, 2}} \, g_2(x^\prime,\,Q^2) \, ,
\eea
where the nucleon-pole contribution is included in the integral on the {\it rhs}.
If we further assume that the amplitude $S_2$ converges faster than $1/\nu$ for
$\nu \rightarrow \infty$,
we may write an unsubtracted DR for
the amplitude $\nu \, S_2$, which is even in $\nu$,
\bea
\label{S2dr2}
{\rm{Re}}  \left[\nu \, S_2(\nu,\,Q^2)\right]
\,=\,    2 \alpha_\mathrm{em} M \,{\mathcal{P}} \,  \int_{0}^{\infty} \,d\nu^\prime \,
\frac{1}{\nu^{\prime \, 2} -\nu^2}  \, g_2(x^\prime,\,Q^2) \, .
\eea
By multiplying Eq.~\ref{S2dr} by $\nu$ and subtracting it from Eq.~\ref{S2dr2}, 
one then obtains the  Burkhardt-Cottingham (BC) ``superconvergence sum rule''~\cite{Burkhardt:1970ti},
valid for {\it{any}} value of $Q^2$:
\bea
\label{BC}
\int_{0}^{1}\, g_{2}\,(x,\,Q^2)\, dx  = 0 ,
\eea
provided that the integral converges for $x \rightarrow 0$. 
The upper integration limit in Eq.~\ref{BC} extends to $1$, and thus includes the elastic (i.e. pole) contribution. 
By separating elastic and inelastic parts in the integral of Eq.~\ref{BC}, the BC sum rule can be expressed equivalently:
\bea
\label{BC2}
 \Gamma_2^{(1) \, inel}(Q^2) \,=\,
\frac{\nu_B}{4 M} \, F_P(Q^2) G_M(Q^2)  .
\eea
In perturbative QCD, the BC sum rule
was verified for a quark target to first order in $\alpha_s$~\cite{Altarelli:1994dj}. 
In the non-perturbative domain of low $Q^2$, the BC sum rule was also verified within HBChPT~\cite{Kao:2002cp,Kao:2003jd}. 

The LEX of the non-pole part of the amplitude $(\nu S_2)$ in Eq.~\ref{S2dr2} can be expressed as~\cite{Drechsel:2002ar}: 
\bea
\label{S2lexnu}
\left[\nu \, S_2(\nu,\,Q^2)\right]^{np} =
\frac{2    \alpha_\mathrm{em} M}{\nu_B} \, \Gamma_2^{(1) \, inel}(Q^2)
\,+\, \left[ \frac{2  \alpha_\mathrm{em} M}{\nu_B^3}  \Gamma_2^{(3) \, inel}(Q^2) \right]  \, \nu^2
\,+\, {\mathcal{O}}(\nu^4) \,.
\eea
The third moments of the spin structure functions $g_1$ and $g_2$ can be combined by defining a  longitudinal-transverse polarizabilitiy $\delta_{LT}(Q^2)$ as~\cite{Drechsel:2002ar}:
\bea
\delta_{LT}\,(Q^2) 
&\;\equiv\;& \frac{2 \, \alpha_\mathrm{em}}{M \nu_B^3}\,
\left\{ \Gamma_1^{(3) \, inel}(Q^2)  + \Gamma_2^{(3) \, inel}(Q^2)  \right\} .
\label{deltaLT} 
\eea
By Taylor expanding $(\nu S_2)^{np}$ in $Q^2$, a further sum rule was obtained in~\cite{Pascalutsa:2014zna, Lensky:2017dlc} for the term in Eq.~\ref{S2lexnu} proportional to $\nu^2$ at $Q^2 = 0$, yielding the polarizability $\delta_{LT}(0)$ as:
\bea
\delta_{LT}(0) =  
- \gamma_{E1 E1} 
+ 3 M \alpha_\mathrm{em} \, \left[ P^{\prime \, (M1, M1)1}(0) -  P^{\prime \, (L1, L1)1} (0) \right]. 
\label{S2sr3}
\eea
Note that similar to their counterpart of Eq.~\ref{qsqrgdhsr}, all quantities which enter Eq.~\ref{S2sr3} are observables in RCS or VCS, 
therefore providing a model independent and predictive relation among low-energy spin structure constants of the nucleon. 
\begin{figure}[h]
\includegraphics[angle=0,width=0.5 \textwidth]{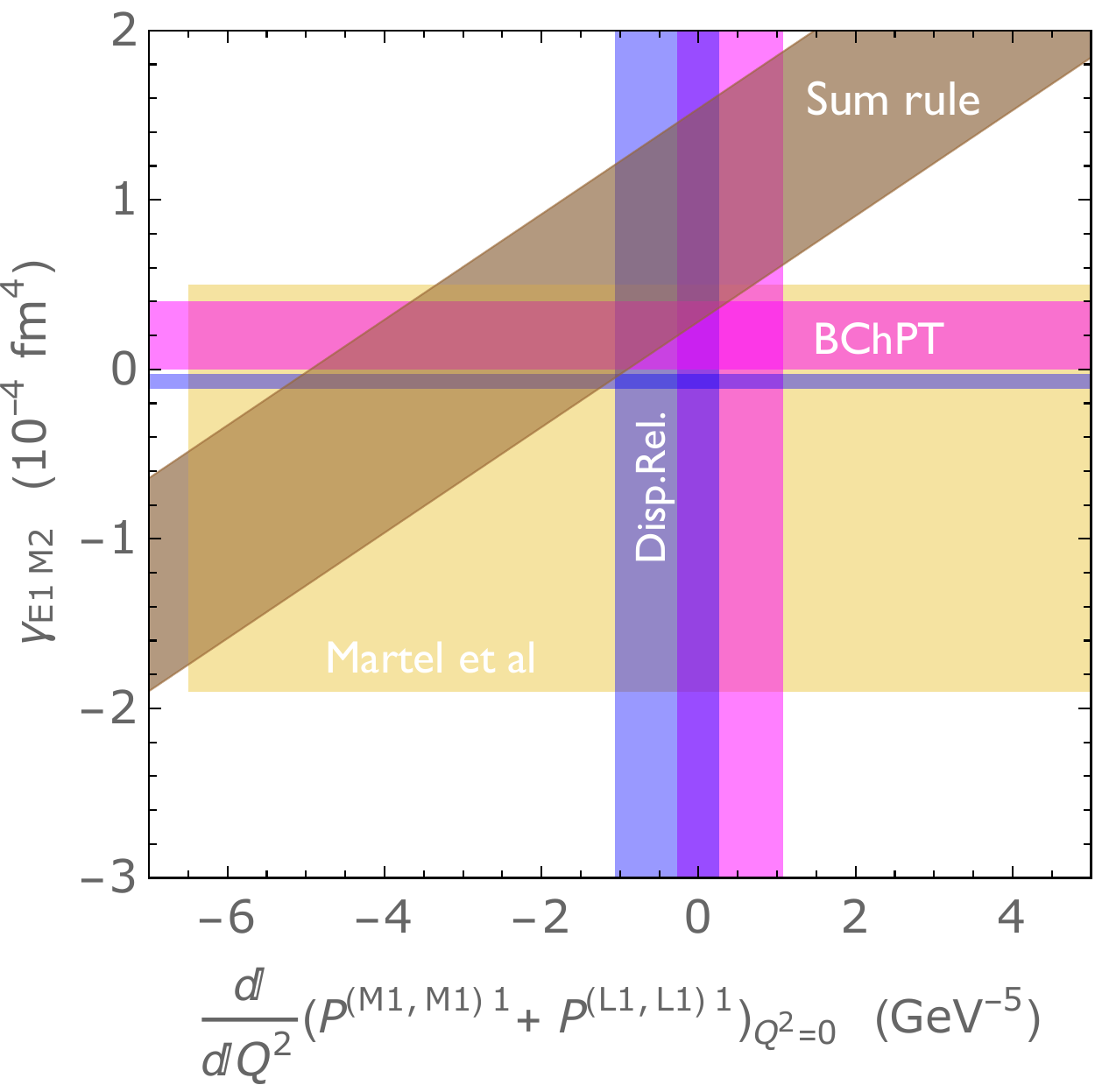}
\includegraphics[angle=0,width=0.5 \textwidth]{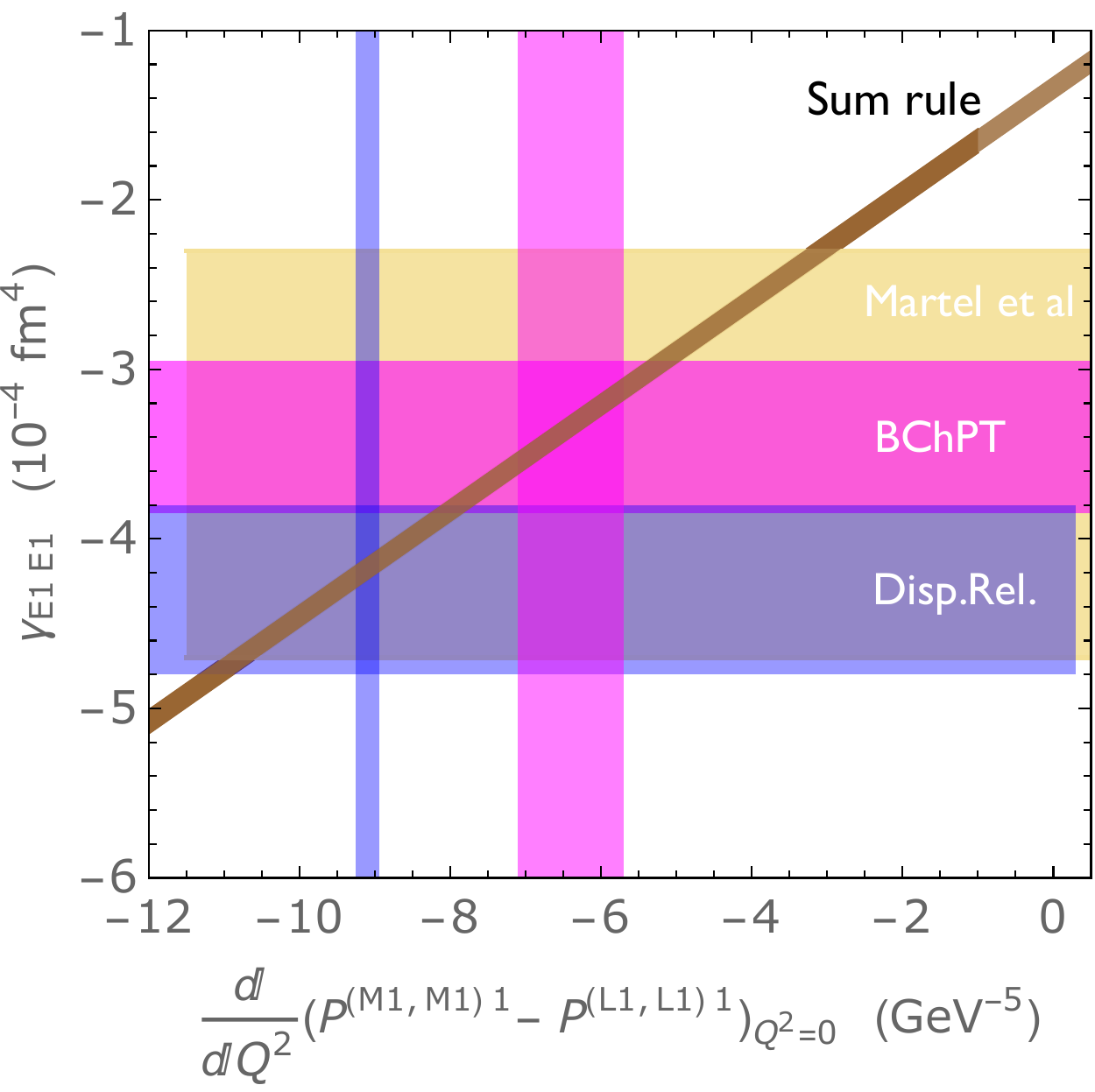}
\vspace{0.1 cm}
\caption{The sum rules of Eq.~\ref{qsqrgdhsr} (left panel) and Eq.~\ref{S2sr3}  (right panel) between proton spin polarizabilities in RCS and VCS respectively. Left: the brown band is the sum rule constraint based on the empirical information for $I_1^\prime(0)$ from JLab/CLAS~\cite{Prok:2008ev,Fersch:2017qrq}, and for $ \langle r_2^2 \rangle$ 
from~\cite{Bernauer:2013tpr}. Right: the brown band is the sum rule constraint based on the phenomenological MAID2007~\cite{Drechsel:2007if}  information for $\delta_{LT}(0)$. 
The yellow bands are the empirical extraction from~\cite{Martel:2014pba}. 
The purple bands are the DR evaluations~\cite{Drechsel:2002ar} for the RCS and VCS polarizabilities, where the width of the bands is obtained by using either 
MAID2000~\cite{Drechsel:1998hk} or MAID2007~\cite{Drechsel:2007if} as input in the dispersive evaluations. 
The pink bands are the BChPT evaluations  
($\pi N + \Delta + \pi \Delta$)~\cite{Lensky:2014dda,Lensky:2015awa,Lensky:2016nui}. Figure from Ref.~\cite{Lensky:2017dlc}.}
\label{Fig:6}
\end{figure}
We provide a graphical presentation of the spin dependent sum rules of Eqs.~\ref{qsqrgdhsr} and \ref{S2sr3} in 
\textbf{Figure \ref{Fig:6}}. 
Using only the empirical information for $I_1^\prime(0)$ and $\delta_{LT}(0)$, the sum rules yield a slanted (brown) band in the plots of $\gamma_{E1M2}$ and  $\gamma_{E1E1}$  
versus the slopes of the GPs.
The pioneering experimental values for $\gamma$'s, obtained by the A2 Coll. at MAMI~\cite{Martel:2014pba}, are shown by the broad horizontal (yellow) bands. The region where the two bands overlap gives a prediction for the slopes of the GPs. A measurement of GP slopes using VCS is required to directly verify this prediction. 
One furthermore sees that the phenomenological DR estimates 
of Ref.~\cite{Drechsel:2002ar} as well as the 
results obtained in  BChPT
are well in agreement, within uncertainties, with the RCS spin polarizabilities and are consistent with the sum rule bands. 
The BChPT results for the slopes of the two spin GPs are also in relatively good agreement with the DR
estimates, as noted in Ref.~\cite{Lensky:2016nui}.

Furthermore, ChPT calculations allow to make predictions for the low $Q^2$ dependence of different moments of spin structure functions as given e.g. in Eqs.~\ref{I1sr} or \ref{deltaLT}. Besides early calculations within 
HBChPT~\cite{Ji:1999pd,Kao:2002cp,Kao:2003jd},  
in more recent years two variants of BChPT have been developed, yielding predictions for different spin structure function moments at low $Q^2$~\cite{Bernard:2012hb,Lensky:2014dda}. 
Although the earlier data~\cite{Kuhn:2008sy,Chen:2010qc} only covered the larger $Q^2$ range, where the theory may lose its predictive power, very recent data~\cite{Fersch:2017qrq,Adhikari:2017wox} and data currently under analysis~\cite{KANG:2013lva,Zielinski:2017gwp} will allow to quantitatively test the predictions of ChPT for the moments of proton and neutron spin structure functions for $Q^2$ down to  $0.01$~GeV$^2$.

\subsection{Spin independent VVCS sum rules}

We next turn to the spin-independent  VVCS amplitudes entering Eq.~\ref{vvcs}. 
Their non-Born parts, denoted by $T^{NB}_1$ and $T^{NB}_2$,  were recently expressed, 
including all terms up to fourth order in $k=\{Q,\nu\}$,  as~\cite{Lensky:2017bwi}:
\bea
T^{NB}_1(\nu, Q^2) &=& 
Q^2 \, \beta_{M1} + \nu^2 \, (\alpha_{E1} + \beta_{M1})  
+ \nu^4 \left[  \alpha_{E1\nu} + \beta_{M1\nu} + \frac{1}{12} (\alpha_{E2} + \beta_{M2})  
\right] \nonumber \\
&+& Q^2 \nu^2 \left[ \beta_{M1 \nu} + \frac{1}{12} (4 \beta_{M2} + \alpha_{E2}) 
+ 2 (\alpha^\prime_{E1}(0) + \beta^\prime_{M1}(0))  
- \alpha_\mathrm{em} 8 M^2  b_{4, 1} 
   \right. \nonumber \\
&+& \left. 
 \frac{1}{M} \left(- \delta_{LT}(0) + \gamma_{M1M1} - \gamma_{E1E1} - \gamma_{M1E2} +  \gamma_{E1M2} 
\right)
+ \frac{1}{(2 M)^2} (\alpha_{E1} + \beta_{M1}) 
\right] \nonumber \\
&+& Q^4   \, \left[ \frac{1}{6} \beta_{M2} 
+ 2 \beta^\prime_{M1}(0)
+ \alpha_\mathrm{em} b_{3, 0}
+ \frac{1}{(2 M)^2} \beta_{M1} 
  \right] 
 + {\cal O}(k^6) , 
\label{lexvcs1_exp} \\
T^{NB}_2(\nu, Q^2) &=& Q^2 \, (\alpha_{E1} + \beta_{M1})  
+ Q^2 \nu^2 \left[  \alpha_{E1\nu} + \beta_{M1\nu} + \frac{1}{12} (\alpha_{E2} + \beta_{M2})  \right]   
\nonumber \\
&+& Q^4 \left[ \frac{1}{6} (\alpha_{E2} + \beta_{M2}) 
+ 2 (\alpha^\prime_{E1}(0) + \beta^\prime_{M1}(0))  
-  \alpha_\mathrm{em} \,4 M^2 b_{19, 0}
\right.  \nonumber \\ 
&& \left. 
- \frac{1}{M} \left( \delta_{LT}(0)  + \gamma_{E1E1} + \gamma_{M1E2} \right)  
+ \frac{1}{(2 M)^2} (\alpha_{E1} + \beta_{M1}) 
 \right] 
+  {\cal O}(k^6) . 
\label{lexvcs2_exp}
\eea
We notice that the quadratic terms are fully determined by the proton electric ($\alpha_{E1}$) 
and magnetic ($\beta_{M1}$) dipole polarizabilities. 
The terms of order $\nu^4$ in $T^{NB}_1$ and of order $ Q^2 \nu^2$ in $T^{NB}_2$ 
are also fully determined by the electric and magnetic dispersive and quadrupole polarizabilities 
which are observables in RCS.
The terms of order $ Q^2 \nu^2$ in $T^{NB}_1$ and of order $Q^4$ in $T^{NB}_2$ involve in addition the slopes $\alpha^\prime_{E1}(0)$ and $\beta^\prime_{M1}(0)$ at $Q^2 = 0$ of the electric and magnetic GPs, shown in Fig.~\ref{fig4}, as well as the RCS spin polarizabilities and the longitudinal-transverse spin polarizability $\delta_{LT}(0)$, all of which are also observable quantities either through RCS, VCS, or using moments of spin structure functions. The only unknowns in these $Q^2 \nu^2$ terms arise from the low-energy coefficients $b_{4,1}$ and $b_{19,0}$, as defined through the expansion for the VVCS used in \cite{Lensky:2017bwi}. We will next show that two forward sum rules will allow to also fix these constants. 
The remaining low-energy constant, $b_{3,0}$ appears at order $Q^4$ in $T^{NB}_1$, and is determined from the $Q^2$ dependence  of the subtraction term in this amplitude, as discussed below. 
  
The DR for the non-pole part  $T_1^{np}$  of the VVCS amplitude $T_1$ requires one subtraction, which we take at $\nu = 0$, in order to ensure high-energy convergence: 
\bea
{\rm{Re}}  T_1^{np}(\nu,\,Q^2)\, & = & \,
T_1^{np}(0,\,Q^2) + \frac{2 \alpha_\mathrm{em} \, \nu^2}{M}  \,{\mathcal{P}}\,
\int_{\nu_{thr}}^{\infty}\, d\nu^\prime \, \frac {1}{\nu^\prime (\nu^{\prime \,2} -\nu^2)}\, F_1(x^\prime, Q^2).
\label{eq:T1dr} 
\eea 
The LEX of the non-pole part $T_1^{np}$, at fixed $Q^2$, takes the form~\cite{Drechsel:2002ar}:
\bea
T_1^{np}(\nu,\,Q^2) \, & = &  T_1^{np}(0,\,Q^2)
\,+\, M^{(2)}_{1}(Q^2)  \, \nu^2 \,+\, M^{(4)}_{1}(Q^2)  \, \nu^4 \,+\,{\mathcal{O}}(\nu^6) \,,
\label{eq:T1lex}
\eea
where $M^{(2)}_{1}(Q^2) $ and $M^{(4)}_{1}(Q^2) $ can respectively be expressed through the second and fourth moments of the unpolarized nucleon structure function $F_1$ as:
\bea
M^{(2)}_{1}(Q^2) =  \frac{2  \alpha_\mathrm{em}}{M \nu_B^2} \int_{0}^{x_{thr}} dx^\prime x^\prime F_1(x^\prime, Q^2) ,
\quad 
M^{(4)}_{1}(Q^2) =  \frac{2  \alpha_\mathrm{em}}{M \nu_B^4}  \int_{0}^{x_{thr}} dx^\prime x^{\prime \, 3} F_1(x^\prime, Q^2).
\label{eq:generalbaldin} 
\eea
To connect the LEX of the non-Born part $T_1^{NB}$ of Eq.~\ref{lexvcs1_exp} with Eq.~\ref{eq:T1lex}, 
we also need to account for the difference between the Born and pole parts. 
As the difference between the Born and pole term contributions to $T_1$ is independent of $\nu$ it can be fully absorbed in the subtraction function $T_1(0,Q^2)$, see Ref.~\cite{Lensky:2017bwi} for details.

The $\nu$-dependent terms in the expansion of Eq.~\ref{eq:T1lex} can then all be determined from sum rules in terms of electro-absorption cross sections on a nucleon.  
The terms of order $\nu^2$ ($\nu^4$) in the LEX of Eq.~\ref{lexvcs1_exp} yield at $Q^2 = 0$ respectively the Baldin sum rule  of Eq.~\ref{baldin} and its higher-order generalization of Eq.~\ref{ho-baldin} as:
\bea
M^{(2)}_{1}(0) = \alpha_{E1} + \beta_{M1}, \quad \quad  
M^{(4)}_{1}(0) =  \alpha_{E1, \nu} + \beta_{M1, \nu} + \frac{1}{12} (\alpha_{E2} + \beta_{M2}). 
\label{eq:higherbaldin}  
\eea
The term proportional to $Q^2 \nu^2$ in the LEX of 
Eq.~\ref{lexvcs1_exp}, yields a new sum rule~\cite{Lensky:2017bwi}:
\bea
M^{ (2)\prime}_1(0) &\equiv&  \frac{d M^{(2)}_1}{d Q^2}  \bigg|_{Q^2 =0} 
=
 \beta_{M1, \nu} + \frac{1}{12} (4 \beta_{M2} + \alpha_{E2}) 
+ 2 (\alpha^\prime_{E1}(0) + \beta^\prime_{M1}(0)) 
-  \alpha_\mathrm{em} 8 M^2  b_{4, 1}
  \nonumber \\
&& + \frac{1}{M} \left(- \delta_{LT} + \gamma_{M1M1} - \gamma_{E1E1} - \gamma_{M1E2} +  \gamma_{E1M2} \right) 
+ \frac{1}{(2 M)^2} (\alpha_{E1} + \beta_{M1}). \quad
\label{eq:mp12}
\eea
The structure function moment $M^{(2)}_1(Q^2)$ is an observable which has been measured at JLab/Hall C~\cite{Liang:2004tk}. 
One can then use the measured value on the {\it lhs} of the sum rule of Eq.~\ref{eq:mp12} in order to determine the low-energy coefficient $b_{4, 1}$.

For the amplitude $T_2$, which is even in $\nu$,  one can write an unsubtracted DR in $\nu$~:
\bea
{\rm{Re}}  T_2^{np}(\nu,\,Q^2)\, =  \,
 2 \alpha_\mathrm{em}  \,{\mathcal{P}}\,
 \int_{\nu_{thr}}^{\infty}\, d\nu^\prime \,   \frac{1}{\nu^{\prime \, 2} - \nu^2}  \,  F_2(x^\prime ,\,Q^2)\,.
\label{eq:T2dr}
\eea
For the amplitude $T_2$ there is no difference between the Born and pole contributions, and its LEX expansion can 
be directly read off Eq.~\ref{lexvcs2_exp}, and expressed as:
\bea
T_2^{np}(\nu,\,Q^2) \, & = &  
M^{(2)}_{1}(0)  \, Q^2
+ M^{(4)}_{1}(0) \, Q^2 \nu^2 
+ M^{(1)\prime}_2(0) \, Q^4
+\,{\mathcal{O}}(k^6).
\label{eq:T2lex}
\eea
One recovers from the $Q^2$ term the Baldin sum rule,  
and from the $Q^2 \nu^2$ term the higher-order Baldin sum rule. 
Furthermore, the term of order $\mathcal{O}(Q^4)$ involves the derivative at $Q^2 = 0$ 
of the first moment of the structure function $F_2$, defined as:
\bea
M^{(1)}_2(Q^2)  &=& \frac{\alpha_\mathrm{em}}{M \nu_B^2}  \,\int_{0}^{x_{thr}} dx^\prime  \,F_2(x^\prime,\,Q^2) ,
\label{eq:m21}
\eea 
which satisfies  $M_2^{(1)}(0) = M_1^{(2)}(0)$.
Its derivative at $Q^2 = 0$ can then be obtained from Eq.~\ref{lexvcs2_exp} through the sum rule relation~\cite{Lensky:2017bwi}:
\bea 
M^{(1)\prime}_2(0)  \equiv \frac{d M^{(1)}_2(Q^2 )}{d  Q^2}  \bigg|_{Q^2 =0} 
&=&  \frac{1}{6} (\alpha_{E2} + \beta_{M2}) 
+ 2 (\alpha^\prime_{E1}(0) + \beta^\prime_{M1}(0)) 
-  \alpha_\mathrm{em} \,4 M^2 b_{19, 0}
\nonumber \\
&-& \frac{1}{M} \left( \delta_{LT}  + \gamma_{E1E1} + \gamma_{M1E2} \right)  
+ \frac{1}{(2 M)^2} (\alpha_{E1} + \beta_{M1}) .
\label{eq:mp21}
\eea 
The knowledge of the slope $M^{(1)\prime}_2(0)$, which is an observable, allows to determine the low-energy coefficient 
$b_{19,0}$.

\begin{figure}[t]
\centerline{%
\includegraphics[width = .5 \textwidth]{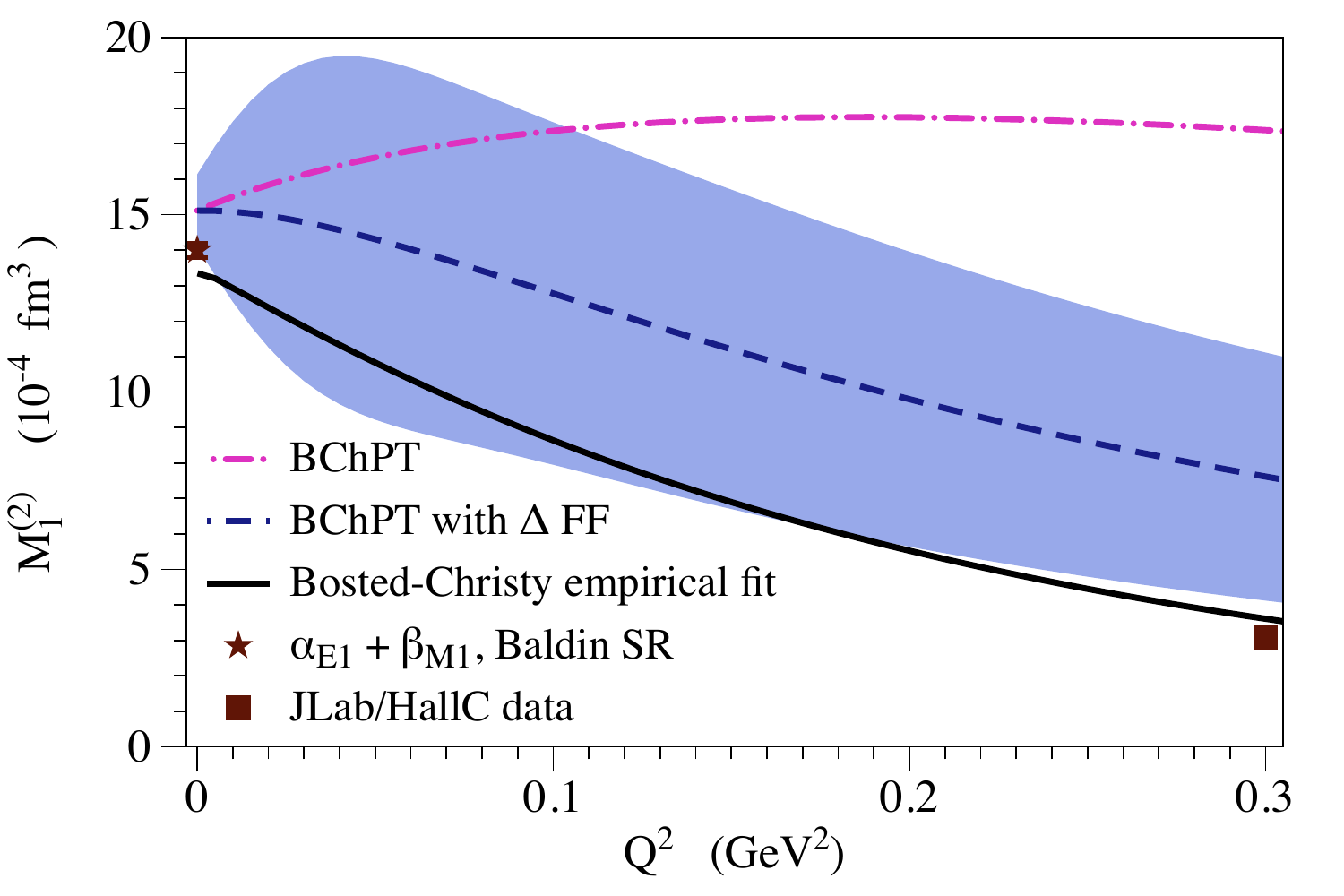}%
\hspace{-6.6cm}
\includegraphics[width = .5 \textwidth]{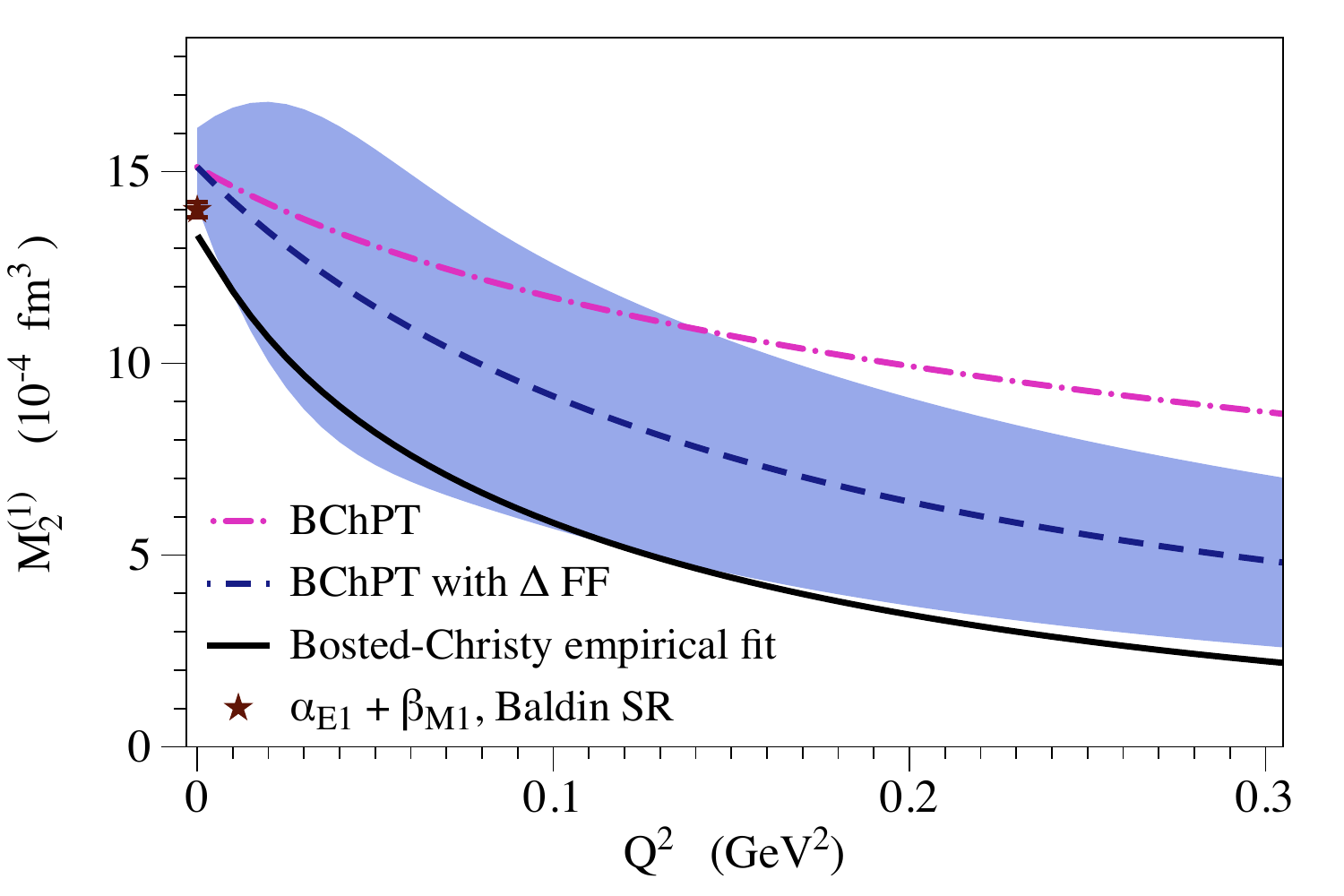}%
}
\vspace{0.12cm}
\caption{$Q^2$ dependence of the proton structure moments 
$M^{(2)}_1$ (left) and $M^{(1)}_2$ (right) according to the empirical Bosted-Christy fit (black solid curve)~\cite{Christy:2007ve}, in comparison with the $\pi N + \Delta + \pi \Delta $ BChPT calculation~\cite{Lensky:2017bwi}. 
For the latter, the blue dashed (magenta dashed-dotted) curves show the results with (without) 
an additional FF dependence in the $\Delta$-exchange, respectively. The blue band shows the uncertainty of the BChPT result with the FF, estimated as in~\cite{Lensky:2016nui}. At the real photon point, both observables yield the Baldin sum rule value for $\alpha_{E1} + \beta_{M1}$~\cite{Gryniuk:2015eza}. 
The data point at $Q^2 = 0.3$~GeV$^2$ (left panel) is from JLab/HallC~\cite{Liang:2004tk}. 
Figure from Ref.~\cite{Lensky:2017bwi}.}
\label{Fig:7}
\end{figure}

\textbf{Figure \ref{Fig:7}} shows the empirical Bosted-Christy fits~\cite{Christy:2007ve} for the moments $M_1^{(2)}$ and $M_2^{(1)}$ in the low-$Q^2$ region, and compares them with the $\pi N + \Delta + \pi \Delta $ BChPT calculation of Ref.~\cite{Lensky:2017bwi}.
One can notice that the BChPT curves agree, within their (rather wide) error
bands, with the empirical fit results.
One can see that the use of FFs
in the $\gamma N\Delta$ vertex is an important part of this result.

\begin{figure}[h]
\begin{center}
\includegraphics[width=0.7\textwidth]{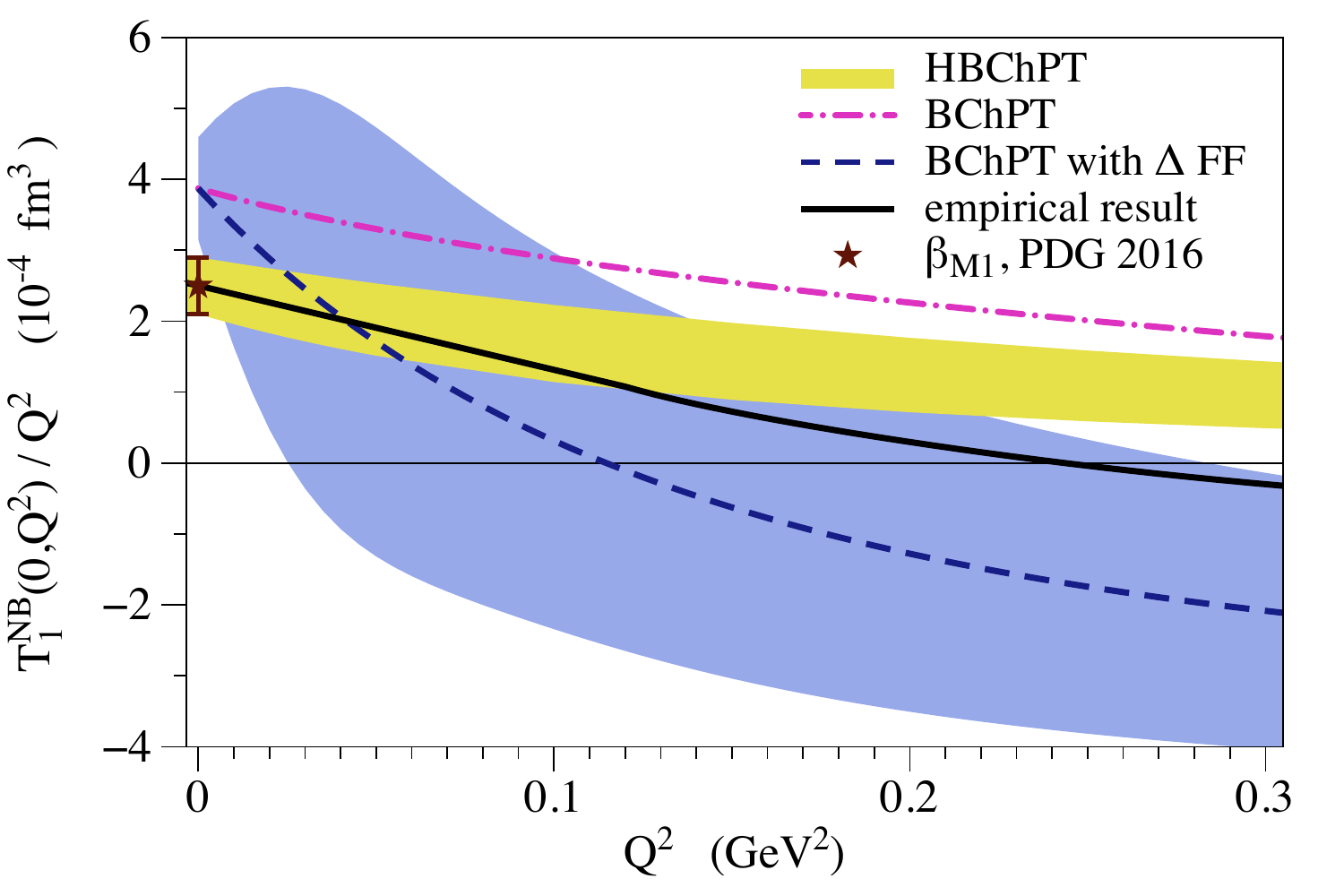}
\vspace{-0.4 cm}
\caption{The low-$Q^2$ behavior of the non-Born part of the subtraction function $T_1^{NB}(0,Q^2)$ divided by $Q^2$. 
The dark yellow band is the HBChPT result~\cite{Birse:2012eb}. 
The blue dashed (magenta dashed-dotted) curves show the BChPT calculation of~\cite{Lensky:2017bwi} with (without)  $\gamma N\Delta$ FF respectively. 
The blue band shows the uncertainty
of the BChPT result with $\gamma N\Delta$ FF, estimated in~\cite{Lensky:2016nui}. The black solid curve shows 
the empirical superconvergence relation estimate of~\cite{Tomalak:2015hva}. 
At the real photon point, the PDG value for $\beta_{M1}$ of Eq.~\ref{alpha-beta-PDG} is shown. Note that the HBChPT curve was shifted to reproduce that value. Figure from Ref.~\cite{Lensky:2017bwi}.
}\label{Fig:8}
\end{center}
\end{figure}

In order to completely fix the term of 
$\mathcal{O}(Q^4)$ in the subtraction function $T_1^{NB}(0,Q^2)$, one needs to 
determine the low-energy coefficient $b_{3, 0}$. Its determination requires a measurement of the VVCS process with a spacelike initial and timelike final photon, which is not available at present.   
As its determination is of importance in the leading hadronic corrections to the proton radius extraction from muonic Lamb shift measurements, we will compare the behavior of $T_1^{NB}(0,Q^2)$ 
in different approaches. \textbf{Figure~\ref{Fig:8}} compares $T^{NB}_1(0,Q^2)/Q^2$ as obtained in
BChPT and HBChPT~\cite{Birse:2012eb}, with a superconvergence relation estimate~\cite{Tomalak:2015hva}. At the real photon point, $T^{NB}_1(0,Q^2)/Q^2$ is given by the magnetic dipole polarizability $\beta_{M1}$. 
The superconvergence and HBChPT estimates were fixed at $Q^2 = 0$ to the 
PDG value for $\beta_{M1}$ of Eq.~\ref{alpha-beta-PDG}, 
whereas the  BChPT estimate reflects the  larger value for $\beta_{M1}$ in this framework.  
One notices from \textbf{Figure~\ref{Fig:8}}  that in the BChPT result, the inclusion of the $\gamma N\Delta$ FF  
yields a suppression at non-zero $Q^2$, yielding a zero crossing in the $Q^2$ range between $0.05 - 0.25$~GeV$^2$. A negative value for $T_1^{NB}(0,Q^2)$ at intermediate $Q^2$ values is also obtained in the empirical superconvergence estimate~\cite{Tomalak:2015hva}. 

A further observable requiring the knowledge of the subtraction function, 
is the electromagnetic mass difference between proton and neutron, 
involving an integral in $Q^2$ of the $T_1^{NB}(0,Q^2)$ difference for proton and neutron, see Refs.~\cite{WalkerLoud:2012bg,Gasser:2015dwa} for some recent works.  

\section{Polarizability corrections to muonic hydrogen spectroscopy}
\label{TPE}

In this section we discuss how the phenomenological information on the unpolarized forward VVCS of the last section is used in  estimates of the two-photon exchange (TPE) corrections to the Lamb shift in muonic atoms. 
The recent extractions of the proton charge radius from the Lamb shift measurements in muonic hydrogen \cite{Pohl:2010zza,Antognini:1900ns} resulted in a significant discrepancy in comparison with measurements with electrons \cite{Bernauer:2010wm,Bernauer:2013tpr,Mohr:2012tt}, see Refs. \cite{Antognini:1900ns,Carlson:2015jba,Hill:2017wzi} for recent reviews. In view of this discrepancy, the higher-order corrections to the Lamb shift were examined in detail by many groups. In particular, the TPE proton structure corrections  were scrutinized over the past decade, see Ref.~\cite{Hagelstein:2015egb} for a review and references therein. The TPE correction contributes at present the largest theoretical uncertainty when extracting the charge radius from the Lamb shift data, thus limiting its accuracy.  In this Section, we briefly review the current status of the dispersive estimates as used in the muonic hydrogen Lamb shift analyses~\cite{Antognini:2013jkc} and compare them with the model independent ChPT analyses. 

The $n$-th $S$-level shift in the (muonic) hydrogen spectrum due to {\it forward} TPE is related to the spin-independent VVCS amplitudes~\cite{Carlson:2011zd} as:
\bea
\Delta E^\mathrm{TPE}(nS)= 8\pi e^2 m \phi_n^2
\frac{1}{i}\int_{-\infty}^\infty \!\frac{d\nu}{2\pi} \int \!\!\frac{d \vec q}{(2\pi)^3}   \frac{\left(Q^2-2\nu^2\right)T_1(\nu,Q^2)-(Q^2+\nu^2)\,T_2(\nu,Q^2)}{Q^4(Q^4-4m^2\nu^2)},
\label{VVCS_LS}
\eea
where $m$ is the lepton mass, $\phi_n^2= (\alpha_\mathrm{em} m_r / n)^3 / \pi$ is the wave function at the origin, with $m_r$  the reduced mass of the lepton-proton system. The polarizability effect on the hydrogen spectrum is described by the non-Born amplitudes $T^{NB}_1$ and $T^{NB}_2$~\footnote{It was correctly remarked in~\cite{Birse:2012eb} that using the conventional definition of polarizabilities in the LEXs of Eq.~\ref{lexvcs1_exp}, implies using the Born term as given by Eq.~\ref{born}. }. 
This polarizability effect can be split into the contribution of the subtraction function 
$T^{NB}_1(0,Q^2)$~\cite{Pachucki:1999zza,Carlson:2011zd}:
\bea
\Delta E^{\mathrm{subtr.}}(nS)= 8 m \alpha_\mathrm{em} \phi_n^2 \,\int_0^\infty \frac{d Q}{Q^3}\frac{v_l+2}{(1+v_l)^2}\, T^{NB}_1(0,Q^2),\label{subpol}
\eea
with $v_l = \sqrt{1+4m^2/Q^2}$, and contributions of the inelastic structure functions~\cite{Hagelstein:2015egb}:
\bea
\Delta E^{\mathrm{inel.}}(nS)&=&-32\alpha_\mathrm{em}^2 Mm\,\phi_n^2\,\int_0^\infty \frac{d Q}{Q^5}\,\int_0^{x_{thr}} d x\frac{1}{(1+v_l)(1+\sqrt{1+x^2\tau^{-1}})} \nonumber \\
&&\times\Bigg\{\left[1+\frac{v_l\sqrt{1+x^2\tau^{-1}}}{v_l+\sqrt{1+x^2\tau^{-1}}}\right]F_2(x,Q^2)\nn\\
&&+\frac{2x}{(1+v_l)(1+\sqrt{1+x^2\tau^{-1}})}\left[2+\frac{3+v_l\sqrt{1+x^2\tau^{-1}}}{v_l+\sqrt{1+x^2\tau^{-1}}}\right]F_1(x,Q^2)\Bigg\}. 
\label{inelasticpol}
\eea
The sum of Eqs.~\ref{subpol} and \ref{inelasticpol} is referred to as the total polarizability contribution. 

{\bf Table \ref{tab3}} shows the TPE corrections due to the inelastic structure functions estimate of~\cite{Carlson:2011zd} and resulting from the subtraction-function estimate of~\cite{Birse:2012eb}, both of which are currently used in estimating the total polarizability contribution to the $2S$-level in the muonic hydrogen analyses~\cite{Antognini:2013jkc}. The estimate of~\cite{Birse:2012eb} assumes a dipole ansatz  for $T_1^{NB}(0,Q^2) / Q^2$, and constrains the mass parameter by a HBChPT calculation to fourth-order in the chiral expansion for the $Q^4$ term in $T_1^{NB}(0,Q^2)$.   
We  compare these results with a LO BChPT analysis,  a NLO BChPT analysis which includes the $\Delta$-pole contribution, and with the NLO HBChPT analysis of~\cite{Peset:2014jxa}. One notices that the BChPT result which includes the $\Delta$-pole is in very good agreement with the DR estimate for the inelastic contribution and with the estimate of~\cite{Birse:2012eb} for the subtraction function contribution. It is also interesting that, although the $\Delta$-pole contributes sizeably to both terms, these contributions come with opposite sign, resulting in a small total polarizability contribution due to the $\Delta$-pole, and a total result close to the LO BChPT estimate.  
In {\bf Table \ref{tab3}}, we also show a NLO HBChPT estimate~\cite{Peset:2014jxa} (last column). Even though it comes with a larger error estimate, its value is larger (in magnitude), deviating by about $2 \sigma$  from the BChPT and DR estimates. It was noticed  however~\cite{Peset:2014jxa}, that the inclusion of the nucleon Born term contributions yields a total TPE result which is similar in size as the DR and BChPT results. 

\begin{table}[h]
\tabcolsep7.5pt
\caption{TPE corrections to the 2$S$-level in muonic hydrogen. All values are given in $\mu$eV. The first two rows are the dispersive ($\Delta E^{\mathrm{inel.}}$) and subtraction function ($\Delta E^{\mathrm{subtr.}}$)  contributions. The sum of both yields the total polarizability contribution ($\Delta E^{\mathrm{pol.}}$). }
\label{tab3}
\begin{center}
\begin{tabular}{@{}l|c|c|c|c@{}}
\hline
 & DR + HBChPT & BChPT (LO) & BChPT (LO + $\Delta$) & HBChPT (NLO)  \\
 & \cite{Carlson:2011zd,Birse:2012eb,Antognini:2013jkc} & \cite{Alarcon:2013cba} & ~\cite{Lensky:2017bwi} & \cite{Peset:2014jxa} \\
\hline
$\Delta E^{\mathrm{inel.}}$  & $-12.7 \pm 0.5$ \quad~\cite{Carlson:2011zd}    & $-5.2$  &  $-11.8$  &$-$  \\
$\Delta E^{\mathrm{subtr.}}$  & \quad  $ 4.2 \pm 1.0$  \quad~\cite{Birse:2012eb}   & $-3.0$  & $4.6$ &$-$  \\
\hline
&&&\\
$\Delta E^{\mathrm{pol.}}$  & \, $-8.5\pm 1.1$ \quad~\cite{Antognini:2013jkc}    & $-8.2^{+1.2}_{-2.5} $  & $-7.2^{+1.2}_{-2.5} $ &$ -26.2 \pm 10.0$ \\
\hline
\end{tabular}
\end{center}
\end{table}

Although the TPE contribution is at present the largest theoretical uncertainty when extracting the proton charge radius from the muonic hydrogen Lamb shift, its total size, as obtained from both DRs and ChPT, is approximately one-tenth as large as would be needed to explain the observed discrepancy in the proton charge radius extraction from electronic or muonic observables, thus leaving the "proton radius puzzle" unresolved~\cite{Carlson:2011zd}. 
The TPE correction is also by far the largest theoretical uncertainty when analyzing the hyperfine splitting in the muonic hydrogen, in this case resulting from the polarized proton structure functions $g_1$ and $g_2$.  We refer  to~\cite{Hagelstein:2015egb} for a recent review of the status of this field.

\section{Conclusions and Outlook}
\label{outlook}

Dispersion relations (DRs) are a powerful tool to  extract information on hadron structure constants from analyses of electromagnetic processes. We have reviewed the real and virtual Compton scattering off the proton  and summarized the recent advances in the  DR analyses applied to such processes. We discussed the latest evaluations of forward real Compton scattering (RCS) sum rules. We furthermore reviewed the application of both unsubtracted and subtracted DR approaches to the non-forward RCS process as tools to extract the proton polarizabilities. The comparison between the fits within DRs and alternative fits within chiral perturbation theories (ChPTs) clearly shows  tension for $\beta_{M1}$, which may depend on the different  choice of the database used in the  analyses. We also discussed recent advances in a multipole analysis of RCS data, along with a recent DR fit of the energy-dependent dynamical polarizabilities. 
We subsequently reviewed the application of DRs to the virtual Compton scattering (VCS) process, and discussed the world data on the generalized polarizabilities (GPs) extracted from such process. Apart from some conflicting data situation in the electric GP around $Q^2 = 0.3$~GeV$^2$ which remains to be sorted out, the VCS world data have allowed to extract the $Q^2$ dependence of the GPs up to values of around $2$~GeV$^2$. We have discussed how these data allow to map out the spatial distribution of the polarization densities in a proton.   
Furthermore, we have reviewed  new sum rules for the forward double-virtual Compton scattering (VVCS)
 process on a nucleon, which allow for model-independent relations between polarizabilities in RCS, VCS, and moments of nucleon structure functions. We have presented the status of these sum rules, using both empirical DR evaluations and  baryon ChPT. Finally, we have  reviewed how this information is  used to predict and constrain the polarizability corrections to muonic hydrogen spectroscopy.

\begin{issues}

We end this review by spelling out a few open issues and challenges 
(both theoretical and experimental) in this field:

\begin{enumerate}
\item
{\it Scalar and spin polarizabilities in RCS}:
Ongoing experiments at MAMI with  transversely and longitudinally polarized targets, and a circularly polarized photon beam, as well as with unpolarized targets and a linearly polarized beam, aim to improve the determination of proton polarizabilities. Much of the data has already been acquired, and the analyses of the polarized data are nearly completed~\cite{Martel:2017pln}. 
A first experiment with a linearly polarized photon beam~\cite{Sokhoyan:2016yrc} has demonstrated a proof of principle for a independent determination of $\beta_{M1}$ from such data. A longer run, with improvements in both the tagging system and the linear beam polarization stability, is underway. This run will provide both a reduction in the asymmetry errors by a factor of about 3.5 and a set of cross-section measurements, that together will enable a separate extraction of $\alpha_{E1}$ and $\beta_{M1}$ at level of the PDG errors~\cite{Martel:2017pln}.
At HI$\gamma$S~\cite{Weller:2009zza}, RCS data has recently been taken on the differential cross section and beam asymmetry at photon lab energy of $85$~MeV at three different scattering angles, and is presently being analyzed.
Further measurements below threshold with transversely polarized target and circularly polarized photon are planned.

\item
{\it Dynamical polarizabilities in RCS}:
The improved statistics and precision of the upcoming data set for the unpolarized cross section and beam asymmetry will definitely help to determine with better accuracy the effects of the leading-order static  and dynamical polarizabilities.
In order to also include  the data above pion threshold in the analysis, a full dispersive treatment for the fit of the dynamical polarizabilities should be developed, by giving up the low-energy expansion of the multipole amplitudes used in the recent analysis of Ref.~\cite{Pasquini:2017ehj}.

\item
{\it Generalized polarizabilities in VCS}:
New data on the unpolarized VCS response functions and GPs have been taken at MAMI and are currently in a final stage of analysis. 
These data will complement the $Q^2=0.2$~GeV${^2}$ points~\cite{Correa} shown in this work.
In particular, expected are data at $Q^2=0.1$~GeV${^2}$ and $Q^2=0.45$~GeV${^2}$,
which are in the domain of applicability of BChPT, and will further test the theoretical predictions.
A newly approved experiment at JLab~\cite{JLab_C12-15-001} which plans to 
measure the unpolarized GPs in the $Q^2$ range of   $0.3-0.75$~GeV$^2$  will be able to shed further light on the conflicting data situation around $Q^2 = 0.3$~GeV$^2$. 
Furthermore, unpolarized VCS data at the same $Q^2$ value for different beam energies allows to separate off the VCS response function labeled $P_{TT}$, which contains only spin GPs.  
This will allow one to experimentally access, for the first time, the dominant spin GP $P^{(M1,M1)1}$ and provide a strong test of the BChPT and DR predictions.   

\item
{\it Further developments of the dispersion formalism}:
As the subtracted DR approach for RCS requires the input from the $t$-channel discontinuities, it can be further improved by a refined analysis of the leading $\gamma \gamma \to \pi \pi$ channel in view of new data for this channel. Furthermore, a better control of the analytical continuation of the $s$-channel contribution into the unphysical region will allow to extend the formalism to energies around the $\Delta$-resonance region at larger value of $-t$ (backward angles). An aim for the VCS process, where currently only an unsubtracted DR formalism has been developed,  is to develop a subtracted DR formalism along the lines of the subtracted DR framework for RCS. To this aim, it will be necessary to have the dispersive input of the $\gamma^\ast \gamma \to \pi \pi$ channel.

\item
{\it Spin structure functions at low $Q^2$}:
Data on the deuteron spin structure function moments of $g_1$ at low $Q^2$, down to $0.02$~GeV$^2$, have recently become available from the JLab/Hall B EG4 experiment~\cite{Adhikari:2017wox}. 
The same experiment also measured the proton spin structure function moments of $g_1$ at low $Q^2$, 
down to $0.01$~GeV$^2$, which are currently under analysis.  
Data on the proton spin structure function $g_2$, using a polarized proton target, are also forthcoming  from the JLab/Hall C 
SANE experiment~\cite{KANG:2013lva} and from the JLab/Hall A g2p experiment~\cite{Zielinski:2017gwp}. In both cases, the data 
analysis is in an advanced stage. Together with the existing data on the neutron spin structure functions, the combined data set will allow for a definitive test of the ChPT results for the low $Q^2$ spin structure function moments.

\item
{\it Subtraction function $T_1(0,Q^2)$ in unpolarized VVCS}: 
Extending the knowledge of this key quantity, which enters both the two-photon exchange (TPE) correction to muonic atoms as well as the electromagnetic mass difference between proton and neutron, beyond the region where ChPT predictions are applicable, is clearly of high interest. 
Superconvergence estimates~\cite{Gasser:2015dwa,Tomalak:2015hva} for the VVCS subtraction function $T_1(0,Q^2$) in the $Q^2 \lesssim 2$~GeV$^2$ region are currently constrained by nucleon structure function data in the resonance region ($W < 3$~GeV) as well as by HERA data at high energies ($W > 10$~GeV). 
However, in the intermediate $W$ region  
($3 \lesssim W \lesssim 10$~GeV), the empirical estimates are quite uncertain due to the scarce data situation in that region. Forthcoming structure function data from the JLab 12 GeV program will allow to further improve such estimates, and extract the VVCS low-energy constant $b_{3,0}$. It may also be very worthwhile to directly access $b_{3,0}$ through a low-energy VVCS experiment, using the $e^- + p \to e^- + p + l^- l^+$ process by measuring a dilepton pair ($l^-l^+$) in the final state. 

\item
{\it Polarizability corrections to muonic atom energy levels}:
Although the theoretical precision of two-photon exchange (TPE) corrections to the muonic hydrogen Lamb shift is currently at the level of the experimental one, the situation still needs to be improved for dispersive TPE estimates for muonic deuterium~\cite{Carlson:2013xea} or muonic $^3He^+$~\cite{Carlson:2016cii} in order to match the experimental precisions. New data on few-body electromagnetic observables from the MESA facility in Mainz~\cite{Denig:2016dqo} hold the promise to provide the required input.  
Furthermore, forthcoming high-precision experiments by the CREMA~\cite{Pohl:2016tqq} and FAMU~\cite{Adamczak:2016pdb} collaborations, and at J-PARC~\cite{Ma:2016etb} aim to measure the 1S hyperfine splitting in muonic hydrogen to a level of precision of $1~\mathrm{ppm}$, exceeding by around two orders of magnitude the current theoretical precision. As the latter is limited by the knowledge of the low $Q^2$ proton's elastic form factors and spin structure functions,
further
studies to improve on their measurements are warranted.

\end{enumerate}
\end{issues}
\section*{DISCLOSURE STATEMENT}
The authors are not aware of any affiliations, memberships, funding, or financial holdings that
might be perceived as affecting the objectivity of this review. 

\section*{ACKNOWLEDGMENTS}
We thank Carl Carlson, H\'el\`ene Fonvieille, Franziska Hagelstein, Vadim Lensky, Vladimir Pascalutsa, Paolo Pedroni, Stefano Sconfietti, and Oleksandr Tomalak for helpful discussions and correspondence, and Vladimir Pascalutsa also for  reading  the manuscript. The work of M.V. was supported by the Deutsche Forschungsgemeinschaft DFG in part through the Collaborative Research Center [The Low-Energy Frontier of the Standard Model (SFB 1044)], and in part through the Cluster of Excellence [Precision Physics, Fundamental Interactions and Structure of Matter (PRISMA)].

%


\newpage

\end{document}